\documentclass[10pt,conference]{IEEEtran}
\usepackage{cite}
\usepackage{amsmath,amssymb,amsfonts}
\usepackage{algorithmic}
\usepackage{graphicx}
\usepackage{textcomp}
\usepackage{xcolor}
\usepackage[hyphens]{url}
\usepackage{array}
\usepackage{dblfloatfix}
\usepackage{placeins}

\ifCLASSOPTIONcompsoc
\usepackage[caption=false, font=normalsize, labelfont=sf, textfont=sf]{subfig}
\else
\usepackage[caption=false, font=footnotesize]{subfig}
\fi

\newcommand{\PageLimit}[1]{{}}

\newcommand{\rv}[1]{{\color{black}#1}}

\def\BibTeX{{\rm B\kern-.05em{\sc i\kern-.025em b}\kern-.08em
    T\kern-.1667em\lower.7ex\hbox{E}\kern-.125emX}}

\def\arraybackslash{\let\\\tabularnewline}

\title{DeFiNES: Enabling Fast Exploration of the Depth-first Scheduling Space for DNN \\Accelerators through Analytical Modeling} 

\newcommand\blfootnote[1]{%
  \begingroup
  \renewcommand\thefootnote{}\footnote{#1}%
  \addtocounter{footnote}{-1}%
  \endgroup
}      
      
\author{
    \IEEEauthorblockN{Linyan~Mei$^\dagger$, Koen~Goetschalckx$^\dagger$, Arne~Symons~and~Marian~Verhelst\\Department of Electrical Engineering - MICAS, KU Leuven, Leuven, Belgium\\Email: \{linyan.mei, koen.goetschalckx, arne.symons, marian.verhelst\}@kuleuven.be}
}

\begin{document}
\maketitle
\thispagestyle{plain}
\pagestyle{plain}


\begin{abstract}

DNN workloads can be scheduled onto DNN accelerators in many different ways: from layer-by-layer scheduling to cross-layer depth-first scheduling (a.k.a. layer fusion, or cascaded execution). This results in a very broad scheduling space, with each schedule leading to varying hardware (HW) costs in terms of energy and latency. To rapidly explore this vast space for a wide variety of hardware architectures, analytical cost models are crucial to estimate scheduling effects on the HW level.
However, state-of-the-art cost models are lacking support for exploring the complete depth-first scheduling space, for instance focusing only on activations while ignoring weights, or modeling only DRAM accesses while overlooking on-chip data movements. These limitations prevent researchers from systematically and accurately understanding the depth-first scheduling space.

After formalizing this design space, this work proposes a unified modeling framework, DeFiNES, for layer-by-layer and depth-first scheduling to fill in the gaps.
DeFiNES enables analytically estimating the hardware cost for possible schedules in terms of both energy and latency, while considering data access at every memory level. This is done for each schedule and HW architecture under study by optimally choosing the active part of the memory hierarchy per unique combination of operand, layer, and feature map tile. 
The hardware costs are estimated, taking into account both data computation and data copy phases. The analytical cost model is validated against measured data from a taped-out depth-first DNN accelerator, DepFiN, showing good modeling accuracy at the end-to-end neural network level. 
A comparison with generalized state-of-the-art demonstrates up to 10$\times$ better solutions found with DeFiNES. 

\end{abstract}
\blfootnote{$^\dagger$These~authors~contributed~equally~to~this~work.}
\section{Introduction \PageLimit{(1)}}
Deep Neural Networks (DNNs) are well established and various kinds of hardware accelerators are being developed to make their execution efficient.
A crucial aspect when developing such accelerators is how the DNNs will be mapped onto them.
Different execution orders can lead to differences in energy, latency and memory footprint that are orders of magnitude large. Therefore, it is important to be able to quickly assess the cost for a given schedule, DNN, and accelerator.
In order to do this, without developing and running compute intensive simulations, an analytical model that supports all kinds of schedules, DNNs, and accelerators is required.

\begin{figure}[!t]
    \centering
    \includegraphics[width=3.36in]{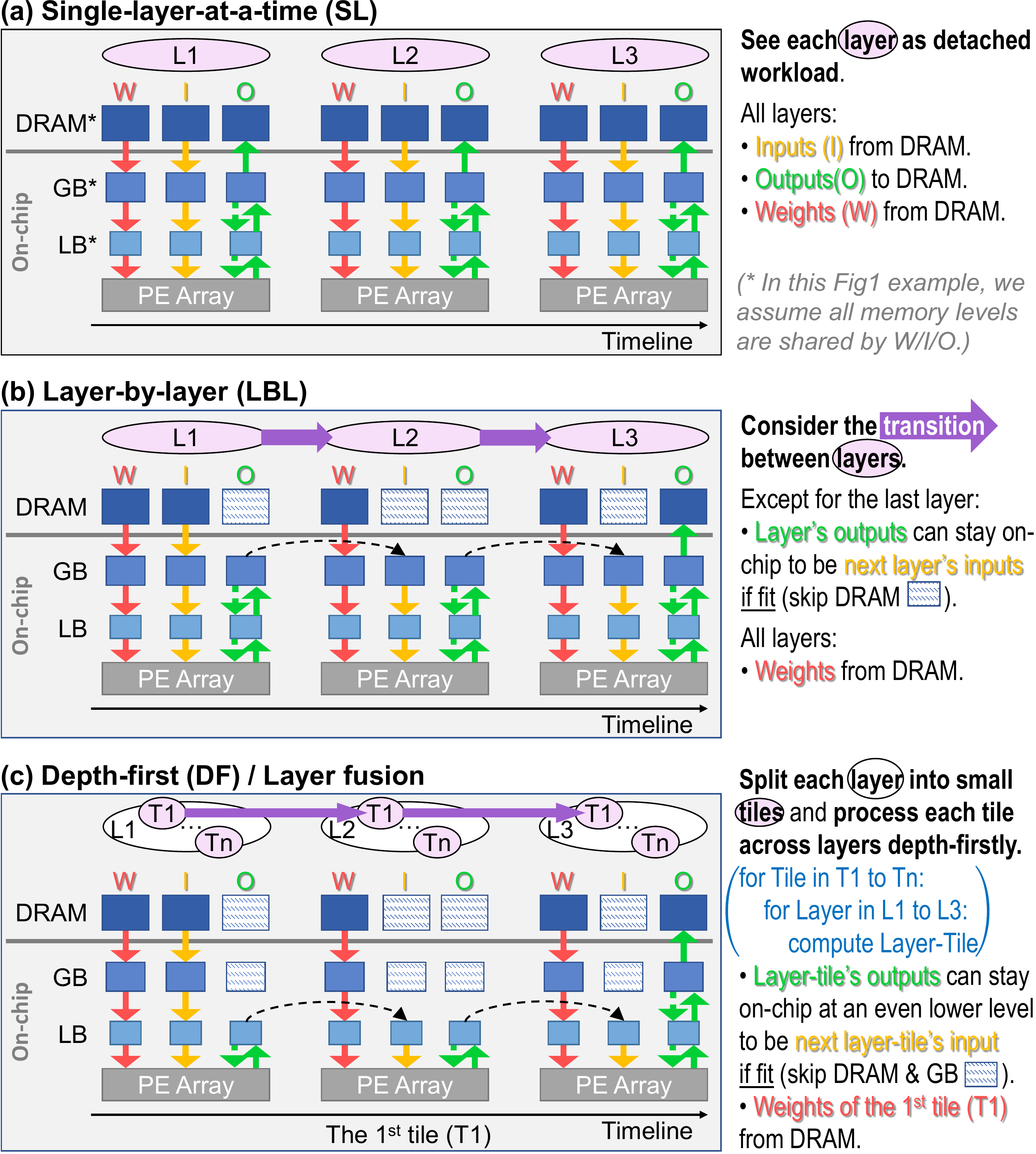}
    \vspace{-0.5em}
    \caption{\rv{Going from (a) Single-layer-at-a-time scheduling to (b) Layer-by-layer scheduling and to (c) Depth-first scheduling to keep activations in lower memory levels. ``L": neural network Layer; ``T": Tile; ``LB": Local buffer (small on-chip memory); ``GB": Global Buffer (larger on-chip memory).}}
    \vspace{-2em}
    \label{fig:intro}
\end{figure}

Analytical models have already been developed to predict the performance of a single layer of a DNN running on an accelerator \cite{MEASTRO, Accelergy, ZigZag}.
However, these ignore cross-layer scheduling possibilities, which can lead to very sub-optimal DNN-level solutions because passing data between layers can have a big impact on the overall system performance~\cite{ConvFusion,Optimus}. A high level example is shown in Fig.~\ref{fig:intro}. In subfigure~(a), representing single layer scheduling, intermediate feature maps are always written to and read from the highest memory level.
We call this `\textit{Single-Layer}' (SL) in this paper.
However, if the feature maps are small enough, it is possible to keep them in lower, more efficient, memory levels, as shown in subfigure~(b).
This is dubbed `\textit{Layer-By-Layer}' (LBL) in this paper.
Furthermore, if the feature maps are too big for this optimization, one can explore \textit{`Depth-First}-like' (DF \cite{df}; a.k.a. layer fusion \cite{15FusedLayerCNN_2016MICRO}, or cascaded execution \cite{TVM_cascade}) scheduling, which means only parts of the intermediate feature maps instead of the whole feature maps are computed at a time and passed between layers.
This decreases the size of data to be passed between layers in a single transaction, which in turn enables the use of an even smaller and more efficient memory level to pass this data (subfigure (c)).

Some accelerators \cite{15FusedLayerCNN_2016MICRO,18AFullHD60_2019VLSI,22DaduEye_2021JSSCC,23A121T_2022JSSCC,24DTCNN_2020TCASI,DepFiN_2021VLSI} already used some forms of such DF scheduling.
However, without a method to quickly explore different DNN accelerators and DF scheduling options, it is hard to say how well these solutions approximate optimality and to quickly estimate the performance of an accelerator in development.

Analytical cost models with support for DF scheduling are thus required to quickly explore and develop optimal systems.
Although such models already exists\cite{DNNFuser,DNNVM,EfficientS,ConvFusion,Optimus}, they are all limited in one or more of the following aspects:
\begin{itemize}
	\item Model only partial hardware cost, like only latency or only DRAM-access, and ignore other relevant costs;
	\item Do not consider an on-chip multi-level memory hierarchy, only distinguish between on-chip and off-chip memory;
	\item Do not study the full DF space (defined in Section \ref{sec:designspace});
	\item Only consider memory accesses for feature map (a.k.a. activation) while ignoring the impact of weights.
\end{itemize}

This work proposes a unified modeling and cost estimation framework, \textit{DeFiNES}\footnote{DeFiNES is open-sourced at \url{https://github.com/ZigZag-Project/DeFiNES}}, for LBL as well as various forms of DF scheduling so as to systematically understand the enlarged scheduling space towards greatly improved energy and latency.
The major highlights of this work are:
\begin{enumerate}
	\item It identifies the full design space of DF scheduling, which also includes SL and LBL by regarding them as two extreme points in the DF design space (Section \ref{sec:designspace});
	\item It presents a Unified Analytical Cost Model that has none of the aforementioned limitations  (Section \ref{sec:model}) and is validated against a taped-out depth-first-style accelerator (Section \ref{sec:validation});
	\item It conducts three case studies based on the model, studying the trade-offs between different DF schedules, and the impact of workload and HW architecture on the \rv{best} DF strategy (Section \ref{sec:casestudies});
	\item It compares DeFiNES against SotA frameworks, showing an up to 10$\times$ better results by including the cost of on-chip memory accesses and accesses caused by weights in the exploration (Section \ref{sec:sota}).

\end{enumerate}

\section{Depth-first Design Space Identification \PageLimit{(1.5)}}\label{sec:designspace}

This section describes the DF design space with three axes, using the well understood LBL inference as a starting point.

\begin{figure}[bt]
	\centering
	\includegraphics[width=3.36in]{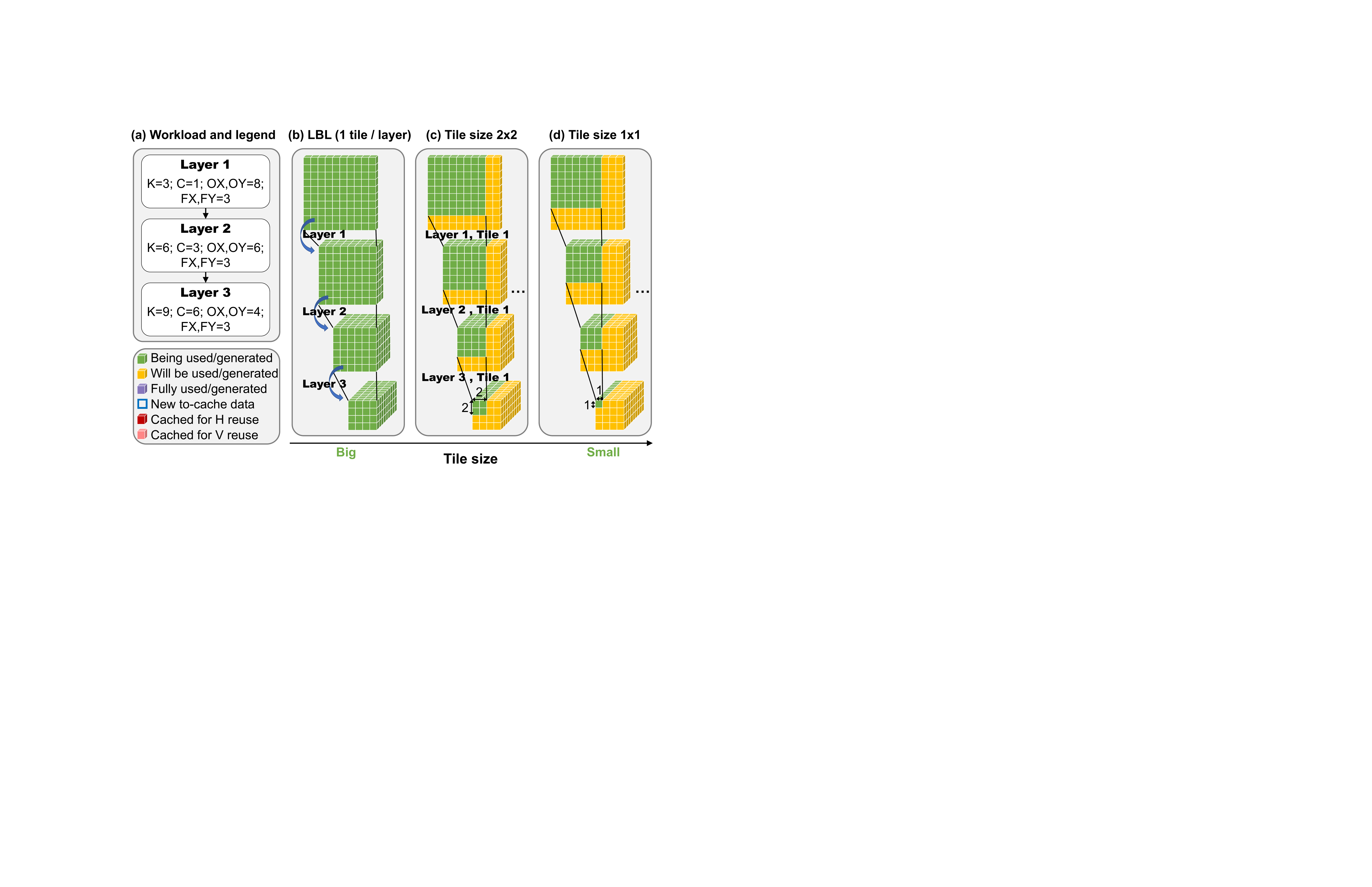}
	\vspace{-0.5em}
	\caption{DF design space's first axis: Tile size. For layer dimension notation in (a): K is for output channel; C is for input channel; OX and OY are feature map spatial dimensions; FX and FY are weight spatial dimensions.}
	\label{fig:tile_size}
\end{figure}

Consider processing multiple layers of a network, as in Fig. \ref{fig:tile_size}(a)\&(b). 
One can calculate the final output feature map in one go, for which the complete input of the last layer is required.
This in turn requires the complete output of the second to last layer, and so on.
Ultimately, this leads to LBL inference, which completely executes each of the layers one at a time starting from the first layer.

Alternatively, one can target to compute only a part of the output of the final feature map.
In this case, only parts of input feature maps are needed, as in shown in Fig. \ref{fig:tile_size}(c).
Inference starts at the first layer, yet only that \textit{tile} of its output feature map that contributes to the target tile in the final output feature map is calculated. 
It is then propagated throughout the other layers to compute the target tile in the final feature map.

\begin{figure}[bt]
	\centering
	\includegraphics[width=3.36in]{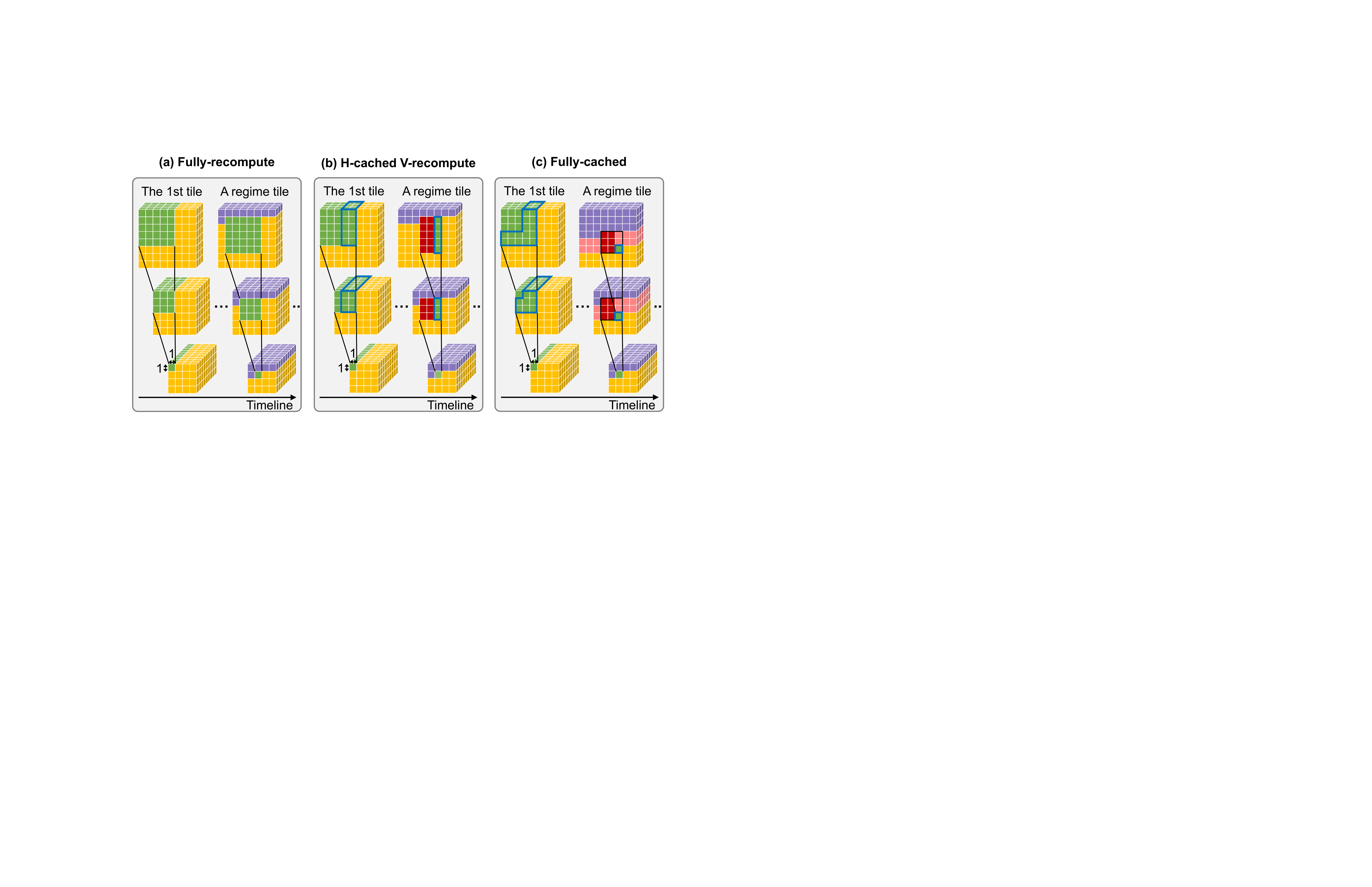}
	\vspace{-0.5em}
	\caption{DF design space's second axis: Overlap storing mode. Workload is Layer 2 and 3 in Fig.~\ref{fig:tile_size}(a); Legend is shared with Fig.~\ref{fig:tile_size}(a).}
	\vspace{-1em}
	\label{fig:mode}
\end{figure}
\begin{figure*}[bt]
	\centering
	\includegraphics[width=6.8in]{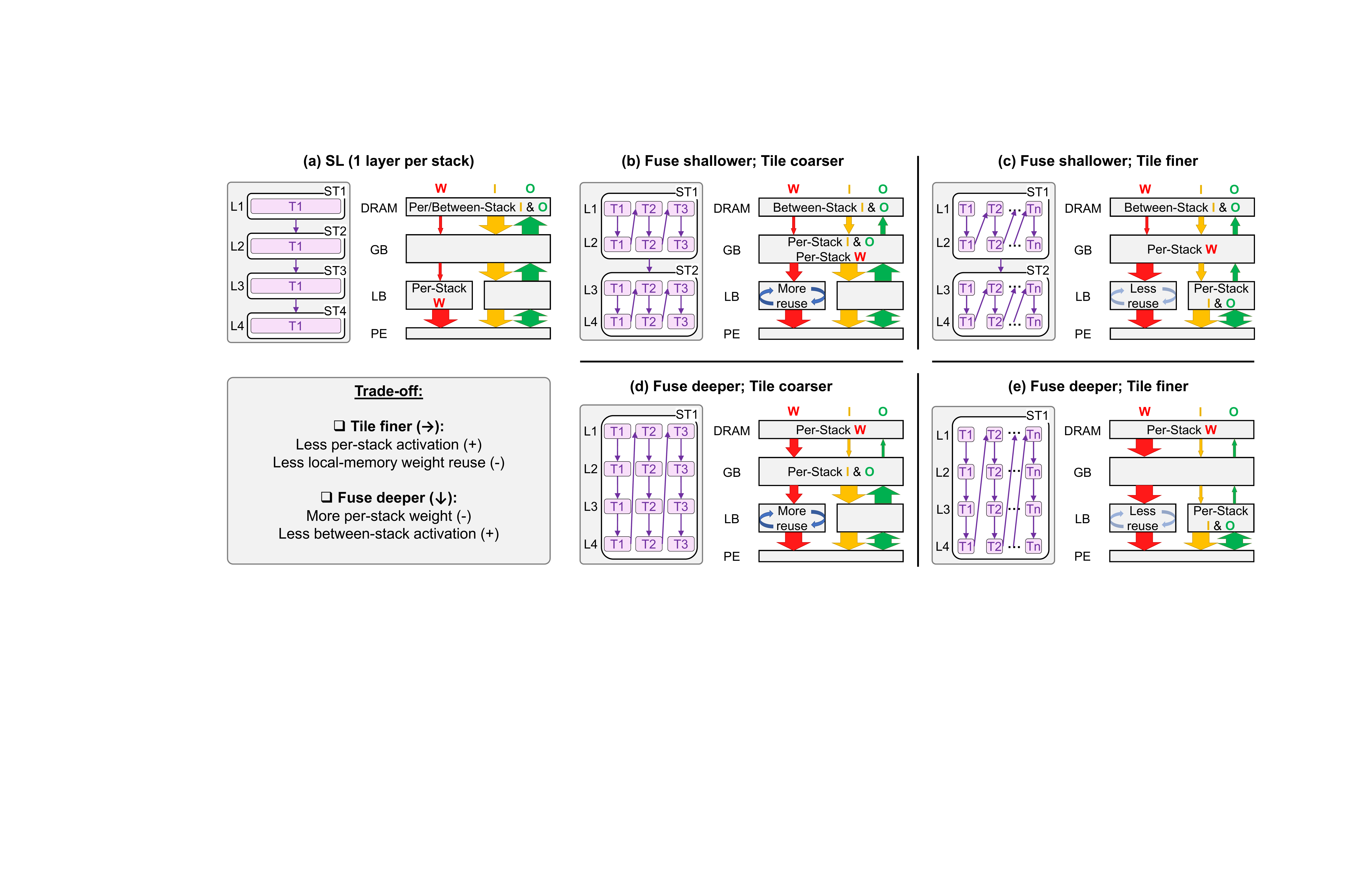}
	\vspace{-0.5em}
	\caption{Impact of tile size (first axis) and fuse depth (third axis). ST: fused-layer STack.}
	\vspace{-1em}
	\label{fig:ST}
\end{figure*}


This illustrates the \textbf{first axis} in the design space: the choice of \textit{tile size}, by which we mean the size of the last layer's portion that we want to compute atomically.
The general trade-off of tile size selection is given in Fig. \ref{fig:ST} (subfigures (b)$\leftrightarrow$(c) or (d)$\leftrightarrow$(e)). 
Choosing a larger/coarser tile size enhances local weight reuse but requires more features to be passed between layers at once, which may require a higher level memory.

Note that in this work, 1) we assume the computation order over tiles is left-to-right, then top-to-bottom and 2) cross-layer tiling is only done across the spatial dimensions (horizontal and vertical dimensions) of the feature maps.
It is not done across the channel dimensions because in most convolution layers all input channels are required to calculate any output feature, which makes cross-layer tiling across the channel dimensions impossible.
However, intra-tile temporal mappings can still have loop tiling over all the dimensions within that tile, including the channel dimensions.
%

Because neighboring tiles of the output feature map can require overlapping parts of earlier feature maps, one can choose either to recompute those overlapped features, or to cache them in some memory in order to reuse them across tiles, as shown in Fig. \ref{fig:mode}.
This choice can be made separately for both spatial dimensions and is considered the \textbf{second axis}.
It has four \textit{modes}: fully-recompute Fig. \ref{fig:mode}(a), horizontally-cached with vertical recompute Fig. \ref{fig:mode}(b), vertically-cached with horizontal recompute, and fully-cached Fig. \ref{fig:mode}(c).
In this work, we don't further consider vertically-cached with horizontal recompute,
\rv{as transposing both the feature maps and, correspondingly, the weights results in the same, yet transposed, outputs, vertically-cached with horizontal recompute and horizontally-cached with vertical recompute are fundamentally the same.}
Choosing caching over recompute requires extra memory space to store the cached data in.
However, it decreases recomputation overhead and the tile size in earlier layers, as Fig. \ref{fig:mode} shows.

So far, this section discussed the scheduling options within one \textit{stack} of \textit{fused} layers.
The final and \textbf{third axis} is the choice of which layers are fused into a stack.
Fusing more layers generally requires more low level weight memory capacity but saves accesses to higher level memories for activations.
This can be seen in Fig. \ref{fig:ST} by comparing subfigures (b) vs. (d), or (c) vs. (e).
Because increasing the memory capacity of the lower level memories decreases their efficiency, the lower level memory can become fruitless if one fuses too many layers.

Note that LBL inference and SL can be positioned in this design space.
On the first axis, the tile size can be set equal to the DNN's final output feature map (Fig. \ref{fig:tile_size}(a)) to get a schedule that is effectively LBL. 
There is only one stack and it executes each layer completely before moving on the next.
One can also choose to only have one layer in every stack (the third axis) (Fig. \ref{fig:ST}(a)), which leads to a SL schedule as we assume features are passed between stacks through the highest memory level.
The second axis has no impact in these cases (LBL and SL) as there is only one tile and thus no overlap between tiles. 

\section{Unified Analytical Cost Model \PageLimit{(3.5)}}\label{sec:model}
\begin{figure*}[h]
	\centering
	\includegraphics[scale=0.97]{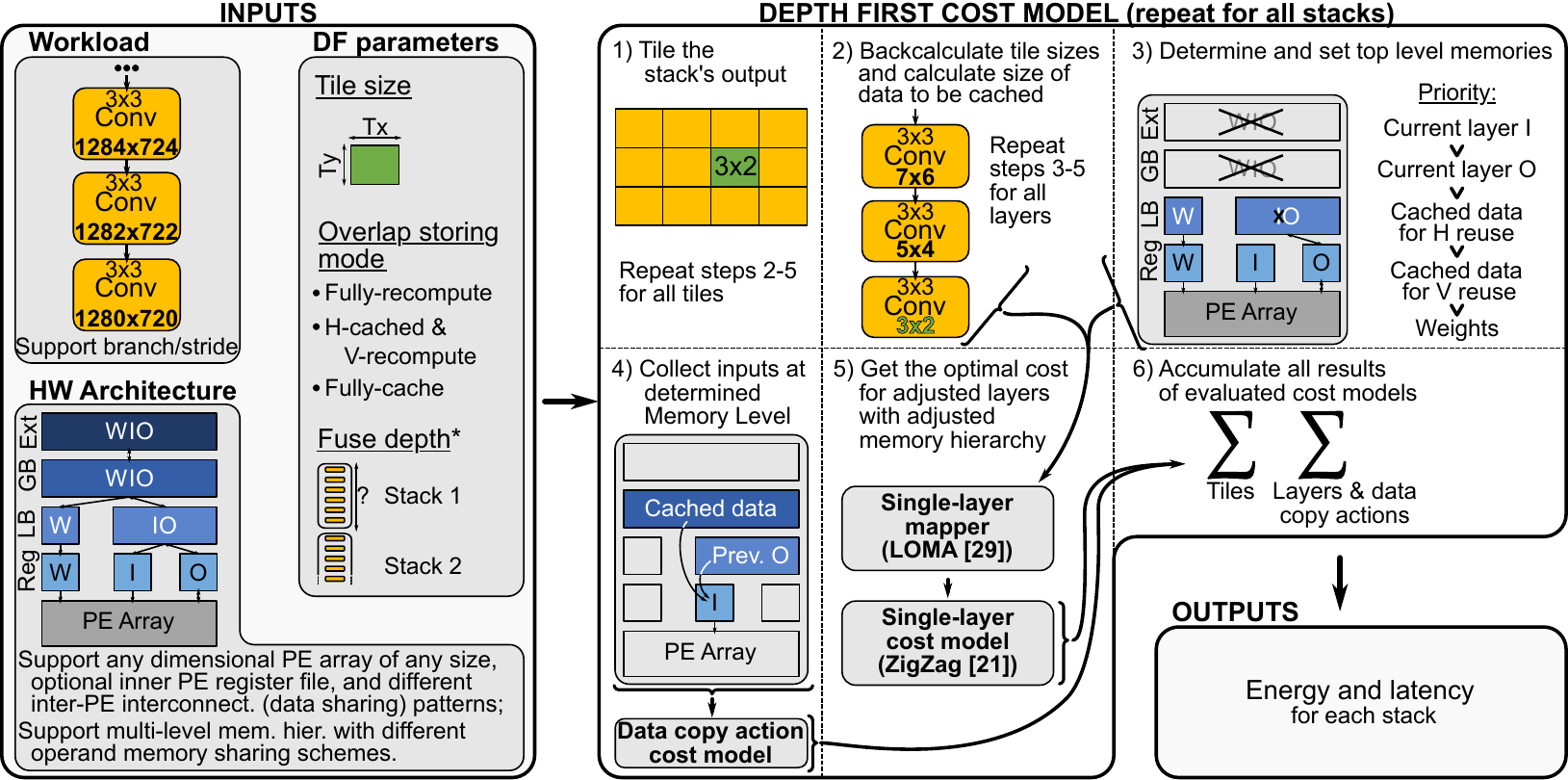}
	\vspace{-0.5em}
	\caption{DeFiNES' overview. (*: optional input, can be set automatically.)}
	\vspace{-1em}
	\label{fig:overview}
\end{figure*}
This section describes the Unified Analytical Cost Model presented in this work, capable of predicting inference costs (energy and latency) of DNNs on a given hardware architecture, with support for the full design space of Section \ref{sec:designspace}.
An overview of the model is depicted in Fig. \ref{fig:overview}.


\rv{The base idea is to use an existing mapping search engine and a cost model that optimize and predict costs for a single layer (step 5 below). However, because of their single-layer limitation, these tools assume every single layer's input and output feature maps need to come from and go to the highest level input and output memories, respectively. 
DeFiNES then provides the Unified Analytical Cost Model as a layer on top of this to provide depth-first compatibility, which it achieves with the following steps:} 



\textbf{Inputs:} The inputs consist of the workload, the HW architecture and the DF parameters.
The workload is a neural network which may have convolution layers, branches, pooling layers, strides, depthwise layers, etc.
The HW architecture consists of an array of Processing Elements (PEs) and a memory hierarchy.
The latter can have memories that are shared between operands (inputs, outputs and weights), different number of levels for different operands, and memories that are unrolled over one or more dimensions of the PE array.
The final input consists of the DF parameters, which identify a point in the design space of Section \ref{sec:designspace}, dubbed the `\textit{DF strategy}'.
The \textit{fuse depth}, i.e. the numbers of layers to fuse together for each stack (third axis), can be given manually or determined automatically.
In the latter case, layers are added to the fused stack as long as the total number of weights in the stack fit in the highest on-chip memory level that holds weights.
In the presence of branching, either all layers between two points where there are no branches are added to a stack, or none of them.
If such a set of layers by itself does not fit in the highest on-chip memory level of weights, none of the layers in this set are fused. 
In other words, each of them is in a 1-layer stack.

With the stacks of fused layers from the workload, hardware, and DF parameters defined, steps 1-6 are done per stack.

\begin{figure}[!t]
	\centering
	\includegraphics[width=3.3in]{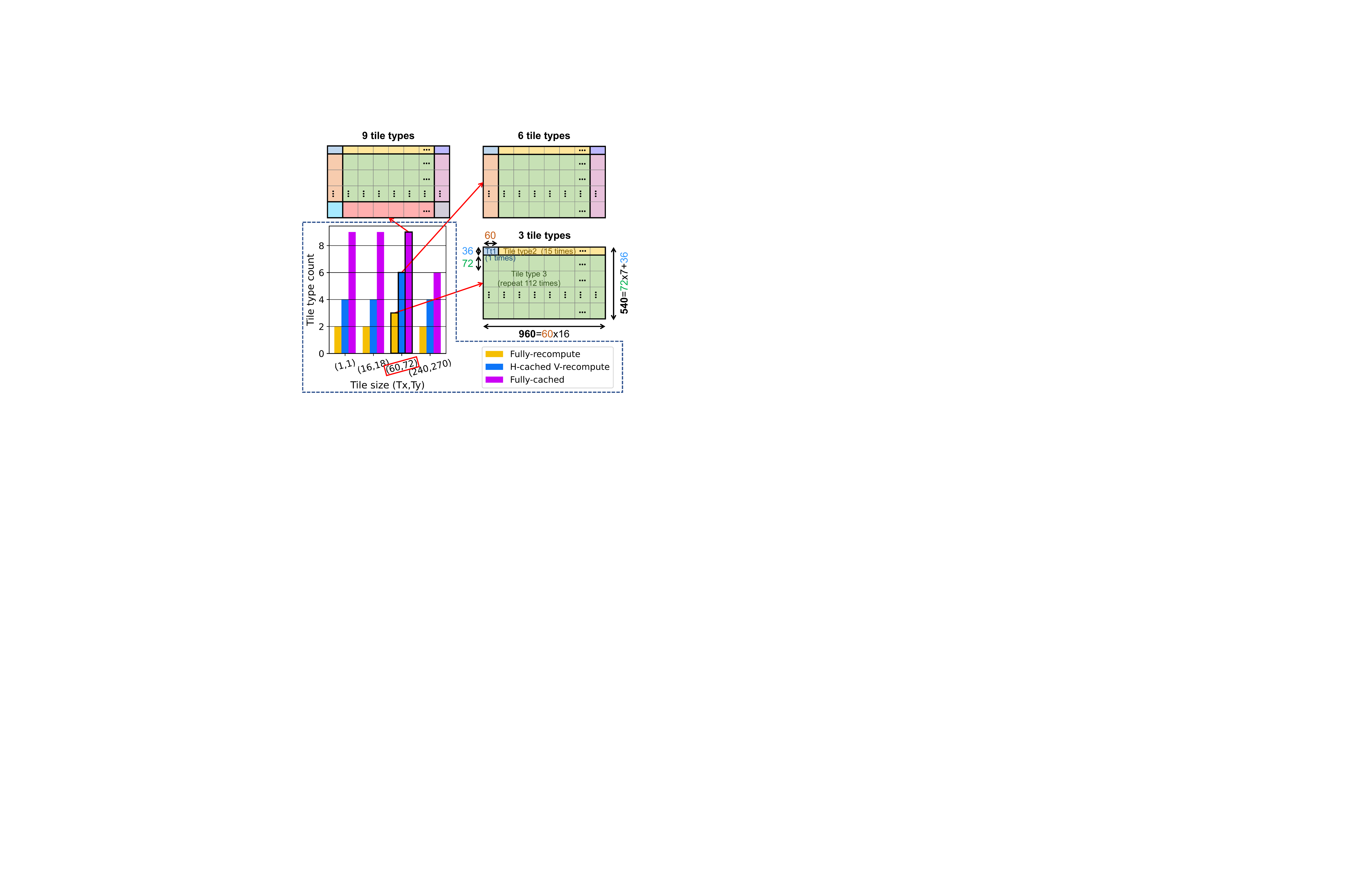}
	\vspace{-0.5em}
	\caption{Tile type count of difference tile sizes and overlap storing modes. The workload used in this example is FSRCNN~\cite{FSRCNN}, whose final output feature map's spatial dimension is 960$\times$540. The 3-tile-type example is further used in Fig.~\ref{fig:tile_memory_levels} and Fig.~\ref{fig:tile_data_sizes}.}
	\vspace{-2em}
	\label{fig:tile_type_count}
\end{figure}

\textbf{1) Tile the stack's output} (for each stack):
Given a stack, the output feature map is partitioned into tiles of the size given by the DF parameters.
As in Fig. \ref{fig:tile_type_count}, the tile size does not have to be a divider of the total feature map size.
Because of this and because tiles in the first row/column do not have cached data available yet -- and similarly the tiles in last column/row do not have to store overlap for their neighbors -- not all tiles are identical.
Therefore, DeFiNES identifies which tiles are completely identical and which are different, a process that leads to different \textit{`tile types'}.
For each tile type, steps 2-6 need to be executed only once as the results can just be replicated for identical copies of the tile, leading to a significant decrease in DeFiNES' runtime.\footnote{\rv{To give a rough idea of how fast DeFiNES (written in Python) runs when we submitted the paper: the Fully-recompute / H-cached V-recompute / Fully-cached with a tile size of (60,72) case of Fig. \ref{fig:tile_type_count} took 23 / 34 / 84 seconds on 1 thread of an Intel Xeon Processor E3-1270 v5, respectively.}}
The number of different tile types also reflects on the code and control complexity of implementing the solution as each tile type can have different set of parameters and temporal mapping, which all need to be programmed into the accelerator.

\textbf{2) \rv{Backcalculate} tile size and calculate the size of data to be cached} (for each tile in each layer):
From the tile size of the last output feature map in the stack, the required tile size of the last layer's input is calculated.
Next, the `to-compute' tile size of the previous layer is calculated.
Without caching for reuse, this simply equals the required tile size of the last layer's input.
However, with caching for reuse across tiles, not all these features need to be calculated as some can be fetched from the cached data, as can be seen in Fig. \ref{fig:mode}(b)\&(c).
This process is repeated for all layers in the stack and as such the input tile size and to-compute output tile size for each layer in the stack is determined.

\begin{figure*}
	\centering
	\includegraphics[width=6.7in]{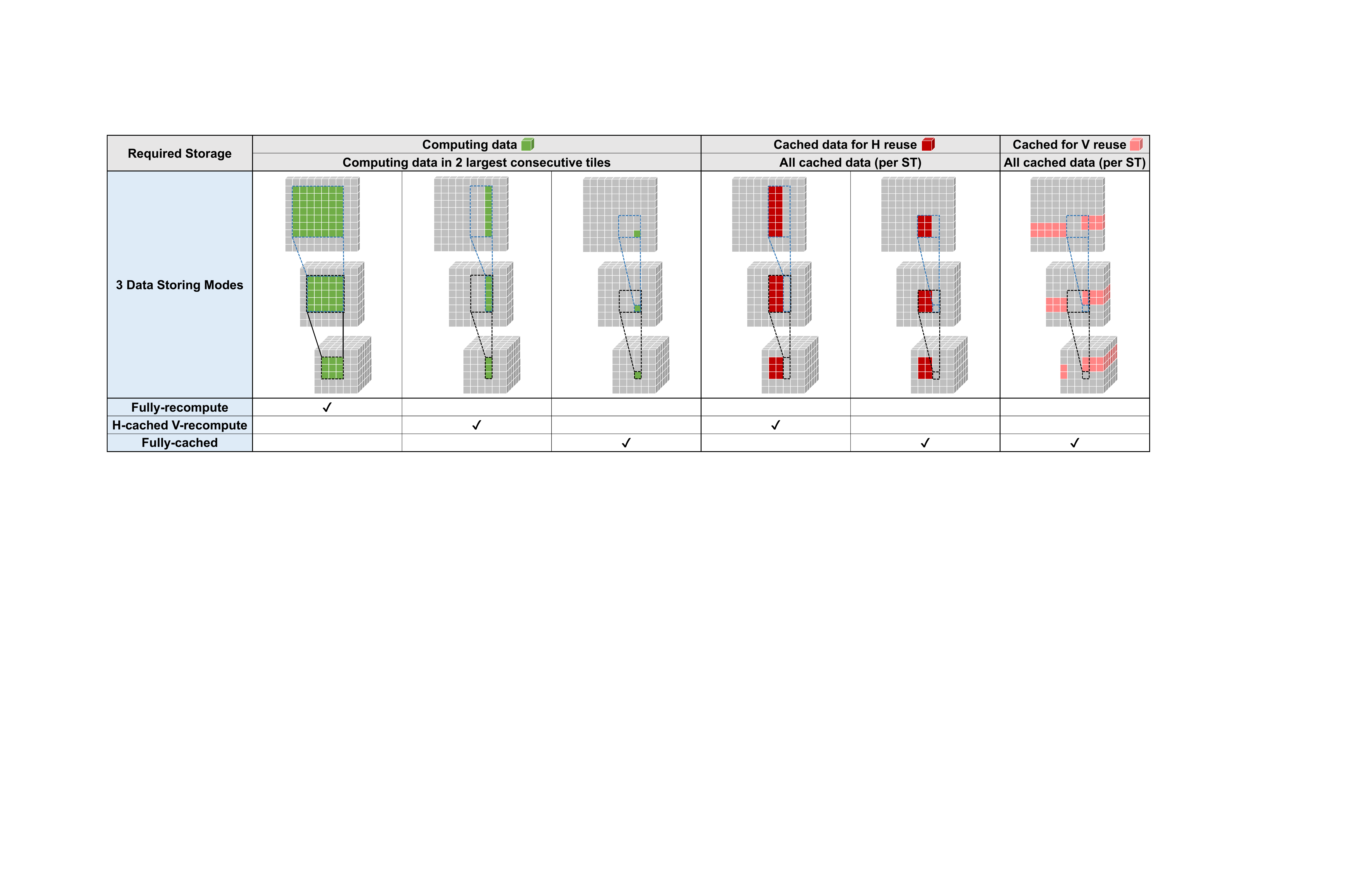}
	\vspace{-0.5em}
	\caption{The required data storage for different overlap storing modes. ST: fused-layer STack.}
	\vspace{-1em}
	\label{fig:Storage}
\end{figure*}

\begin{figure}
	\centering
	\includegraphics[scale=0.80]{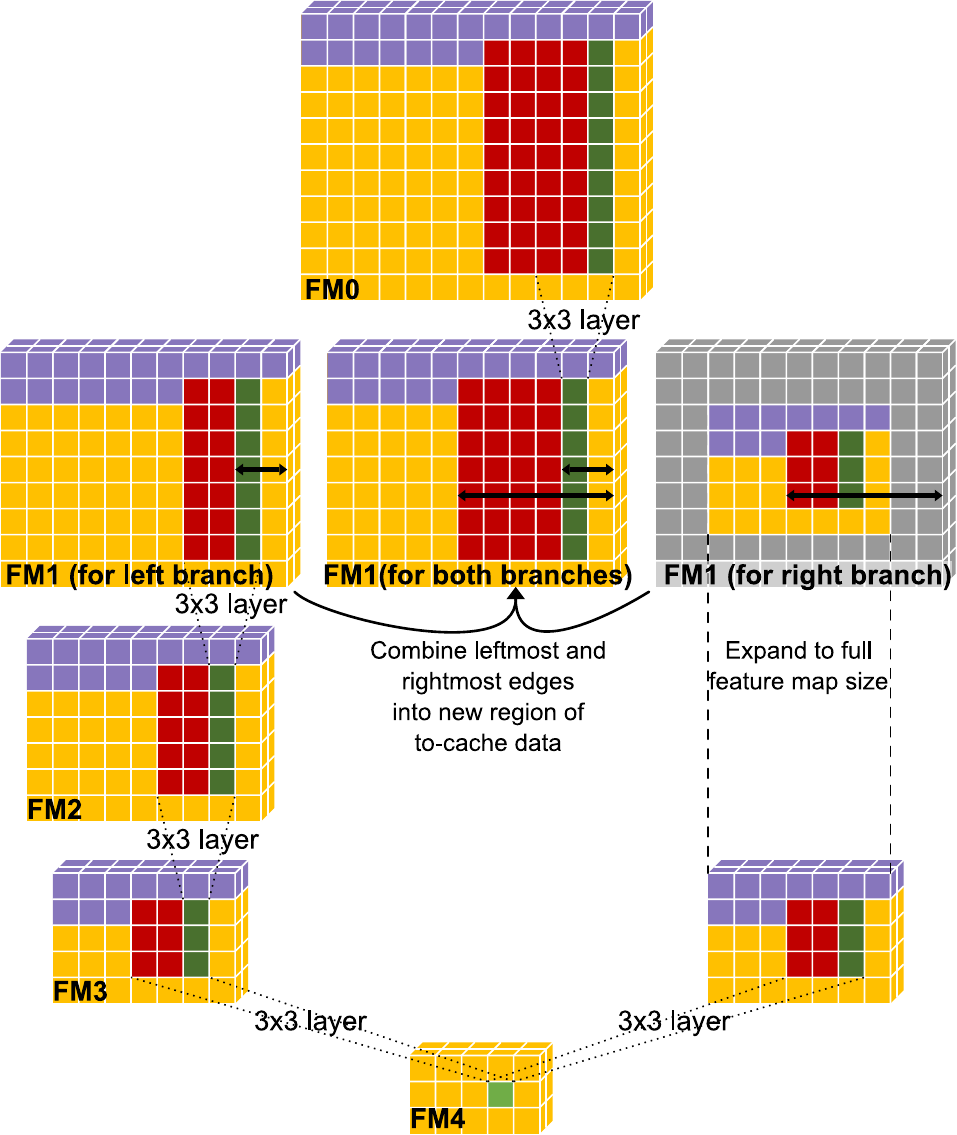}
	\caption{DeFiNES' handling of branching. Legend is shared with Fig.~\ref{fig:tile_size}(a). The grey pixels do not contribute to the right branch. `FM': feature map.}
	\vspace{-1em}
	\label{fig:branching}
\end{figure}

During this process of \rv{\textit{backcalculation}}, the algorithm also keeps track of how much data (from earlier or for future overlapping tiles) of each type in Fig. \ref{fig:Storage} should be cached.
In case of branching, this is handled as in Fig. \ref{fig:branching}. 

In the shown example, the left and right branch need cached features from different places in the feature map.
In such a case, the overall region of features to be cached is set by combining all outermost edges of the to-cache regions, so that all branches always have the cached features they need to operate in overlap caching mode, as can be seen in the middle, combined visualization of FM1 in Fig. \ref{fig:branching}. 

\textbf{3) Determine and set top level memories} (for each tile in each layer):
Given the data sizes calculated in step 2, step 3 determines the highest memory level each type of data (layer inputs, layer outputs, and cached data for H-cached and/or V-cached modes) should be stored in.
In these decisions, data is prioritized as in Fig. \ref{fig:overview}(3), with higher-priority data assigned to the lower, more efficient memory levels.

\begin{figure}[!t]
	\centering
	\includegraphics[width=3.1in]{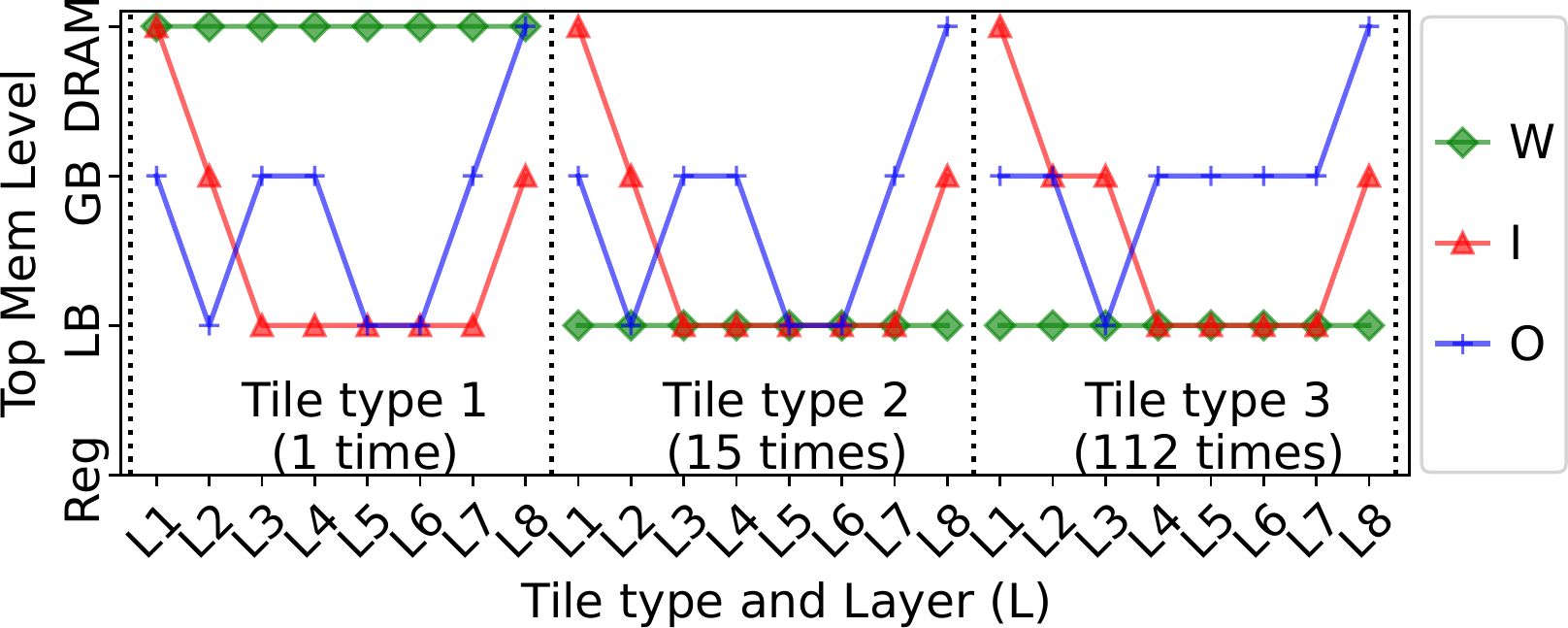}
	\vspace{-0.5em}
	\caption{A visualization of the determined top memory level of each unique layer-tile-combination for operands W, I, and O. The DF schedule is taken from the 3-tile-type example in Fig~\ref{fig:tile_type_count}. The HW architecture is the Idx 2 in Table~\ref{table:CS_setting}. It is worth noting that: 1) for weights, all the layers of the first tile take weights from DRAM, and the other layer-tile-combinations take weights from LB; 2) for input and output, all the tiles' first layer gets input from DRAM, all the tiles' last layer writes output back DRAM, and in between either GB or LB is taken as each of their top memory level.}
	\vspace{-0.5em}
	\label{fig:tile_memory_levels}
\end{figure}

\begin{figure}[!t]
	\centering
	\includegraphics[width=3.1in]{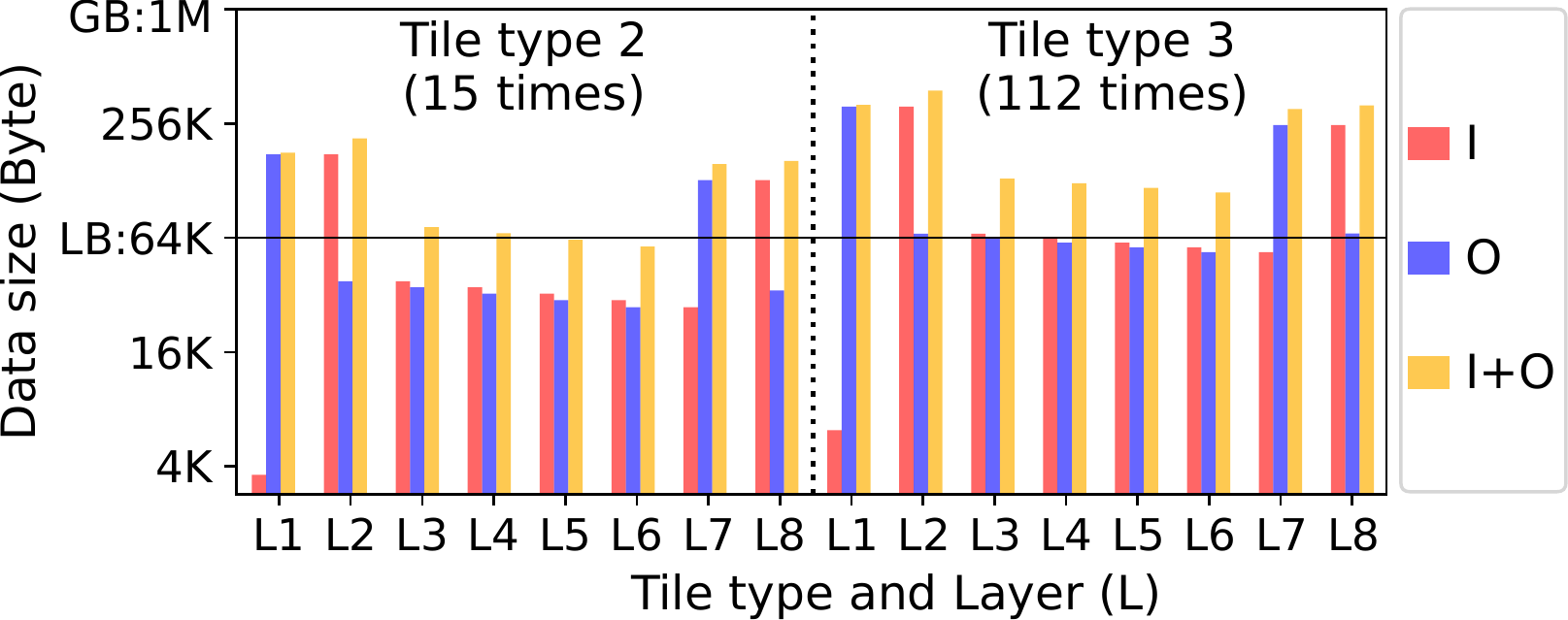}
	\vspace{-0.5em}
	\caption{A visualization of activation data size in tile type 2 and 3 of the example in Fig.~\ref{fig:tile_memory_levels}. The capacities of LB and GB are marked out on y-axis. Fig.~\ref{fig:tile_memory_levels} and Fig.~\ref{fig:tile_data_sizes} together show that 1) when the total activation size (I+O) can fit into LB (e.g., Tile type 2 - L6), the LB is the top memory for both I and O; 2) when the total activation size (I+O) cannot fit into LB while either I or O can fit (e.g., Tile type 3 - L6), I is prioritized to use LB as its top memory level while O is pushed to GB.}
	\vspace{-1em}
	\label{fig:tile_data_sizes}
\end{figure}

Note that the top memory level assigned to different data types can differ between tiles and layers. 
Fig. \ref{fig:tile_memory_levels} gives an example of this for a stack in fully-recompute mode, based on the data sizes from Fig. \ref{fig:tile_data_sizes}.

To be able to use the single-layer mapper and cost model, which assume inputs and outputs come from and go to the top level memory, we remove the initial assignments from the HW architecture's definition of operands to higher memory levels.
We then give this mapper and cost model that adjusted hardware architecture as its input to prevent it from fetching data from or storing data to unnecessarily high memory levels. 

\textbf{4) Collect inputs at determined memory level} (for each tile in each layer):
A single layer-tile-combination can have input feature data that is located in different memory levels, for instance the newly created output of a previous layer can be in a lower memory level than cached data from a previous tile.
Therefore, before calling the single-layer mapper and cost model, we model the action of collecting these data into the single memory level that was decided to serve as the top level memory for inputs in step 3. 

Each such data collecting action is defined as a \textit{data copy action}, which is modelled by its to-move data type and amount, the source memory level, and the destination memory level.
The cost of data copy action is calculated by the \textit{data copy action cost model}. This model takes in 1) a list of data copy actions (those actions can theoretically happen in parallel) and 2) the HW architecture (with all the memory port type, port connection, word-length, and per-word access cost defined) to analyze the energy and latency this bundle of data copy actions costs, taking into account possible memory port conflicts in the concurrent actions.

\textbf{5) Call single-layer mapper and cost model} (for each tile in each layer):
At this point, the single layer temporal mapping search engine and cost model are used to get the cost for a single layer-tile-combination. \rv{For this paper, we used LOMA\cite{LOMA} as the mapping search engine and ZigZag\cite{ZigZag, LatencyModel} to extract the cost. Note that other single-layer mappers and cost models (such as \cite{Timeloop, Accelergy, MEASTRO, Sparseloop, MindMappings, GAMMA} and so on) can also be plugged in DeFiNES to serve the purpose.}

\textbf{6) Accumulate results} (for each stack, across all tiles and layers):
Finally, the results of all cost models evaluated in steps 4 and 5 are summed together to get the final energy and latency cost for the stack.






\section{\rv{Validation} \PageLimit{(1)}}\label{sec:validation}
\rv{To extract costs for a single layer-tile-combination, DeFiNES makes use of the ZigZag framework, which is already well validated against several measured hardware \cite{Eyeriss,TinyVers,Meta_prototype}, as well as other SotA cost model \cite{Accelergy}, for single-layer execution. To also ensure good cost modeling of depth-first/layer-fused execution of complete networks, DeFiNES is validated with end-to-end network.
We validate full network performance predictions of DeFiNES by comparing them against hardware measurements of DepFiN~\cite{DepFiN_2021VLSI}, one of the few existing depth-first neural network processors.
For this comparison, we describe DepFiN's core and memory hierarchy in DeFiNES' terminology and fix the full temporal mapping to match DepFiN's.
We further use three neural networks for the validation:  1) FSRCNN \cite{FSRCNN},  2) MC-CNN fast \cite{mccnn}, and 3) a simple custom reference network that exists of 10 layers of K=32 and Fx=Fy=3 followed by a final layer of K=16 and Fx=Fy=1 and operates on 1280$\times$720$\times$3 inputs.

\begin{figure}[b]
	\centering
	\vspace{-1em}
	\includegraphics[scale=0.95]{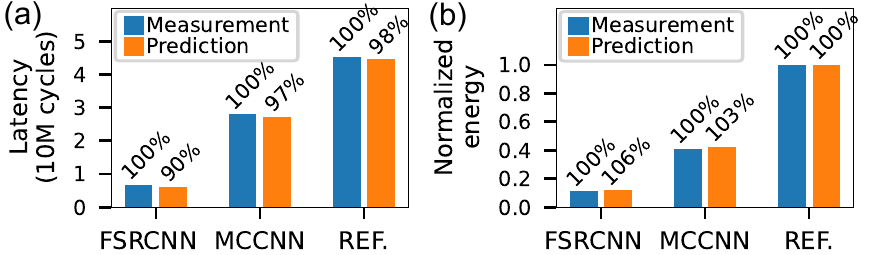}
	\caption{\rv{Compare DeFiNES' results against DepFiN~\cite{DepFiN_2021VLSI} chip measurements.}}\label{fig:valcnn}
\end{figure}

First, Fig. \ref{fig:valcnn} (a) gives the validation results for latency, which shows that DeFiNES' predictions match within 3\% for the second and third network. For the first network, FSRCNN, the error is slightly higher at 10\%. This is because of the stalls caused by
DepFiN's controlling microprocessor that can not fully keep up with the frequent layer switching due to the very small kernels found in FSRCNN. This control flow limitation is not modeled in DeFiNES.

Second, energy is more challenging to match end-to-end, as it is very sensitive to several fine-grain design and layout aspects such as: 1) sparsity, which is used by DepFiN to gate off logic activity to save power; 2) Place-and-Route effects, which cause data transfers to be more expensive than just the memory read/write costs and also includes a sparsity-dependent effect; 3) Process, Voltage, and Temperature (PVT) variations. 
Although these aspects hinder accurately predicting absolute energy consumption, we argue that for the purpose of scheduling optimization it is relative modeling accuracy which matters most in order to be able to choose the best option.
Fig. \ref{fig:valcnn} (b) show the relative energy per inference of the 3 networks, normalized to the reference network inference energy to cancel out the impact of PVT aspects, while aspects 1 and 2 are lumped into the unit cost of the MACs and energy per access of DeFiNES. As can be seen on Fig. \ref{fig:valcnn} (b), the model show to match within 6\% of measurements, building confidence to use DeFiNES for further scheduling optimizations.

}

\section{Case Studies \PageLimit{(2.75)}}\label{sec:casestudies}
Empowered by DeFiNES, three case studies are conducted in order to answer three key DF scheduling questions: CS.1) Given a HW architecture and a DNN workload, how do different DF strategies impact the overall energy and latency? CS.2) Given a HW architecture and multiple DNN workloads (some are activation-dominant while some are weight-dominant), how does the scheduling choice change across different workloads? CS.3) 
Given multiple HW architectures (some are designed for LBL processing while some are manually tuned to be more DF-friendly) and the DNN workloads from CS.2, how do different architectures behave on their optimal scheduling strategies?

\subsection{An overview of experiment settings \PageLimit{(0.25)}}
Table~\ref{table:CS_setting} summarizes the key attributes of different HW architectures and DNN workloads used in the case studies. 

\begin{table*}[t]
	\centering
	\caption{(a) The five HW architectures and their DF-friendly variants and (b) five DNN workloads used in the case studies}
	\includegraphics[width=7in]{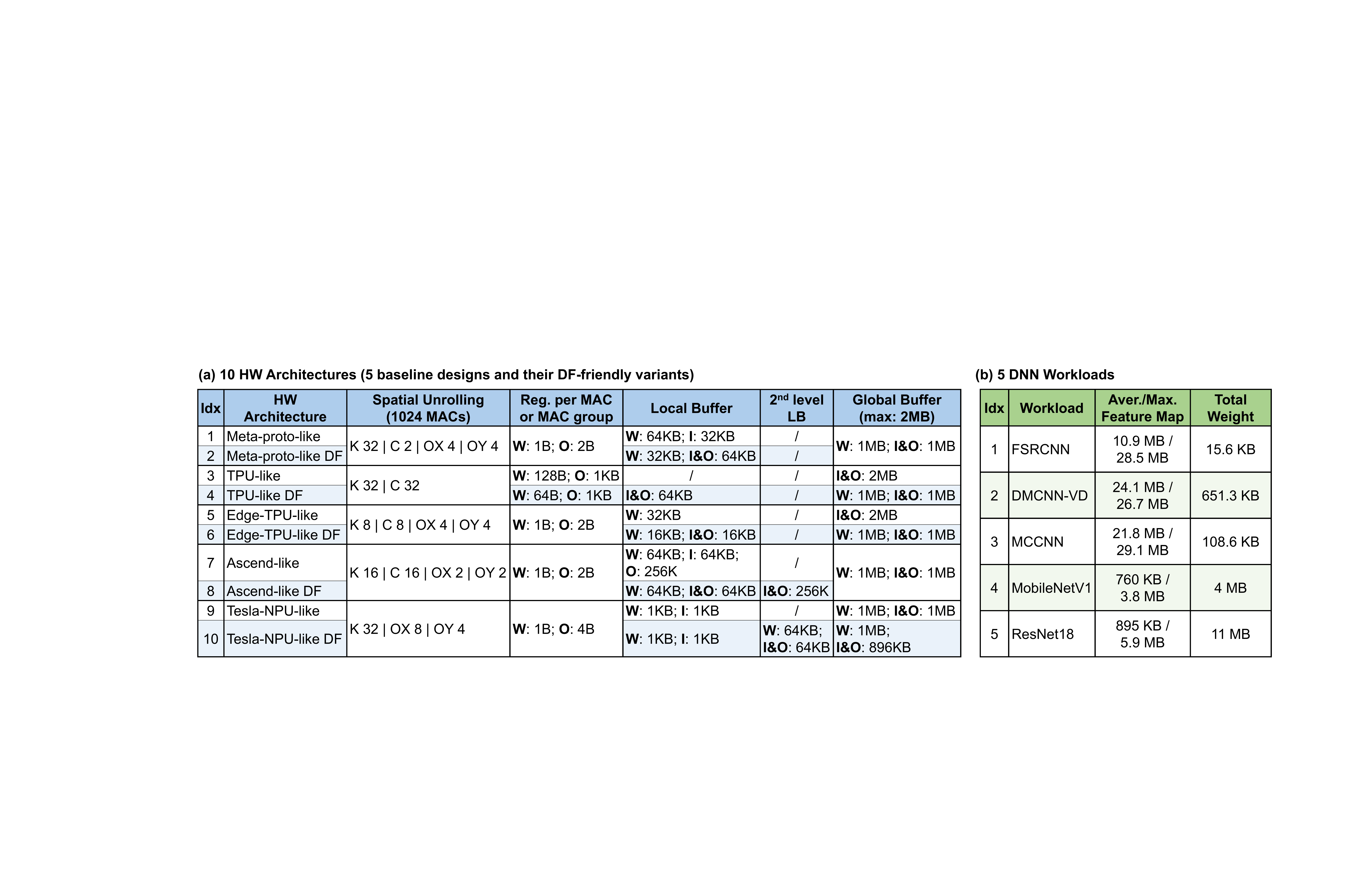}
	\label{table:CS_setting}
\end{table*}

For HW, five DNN accelerators are selected as the architecture baselines for the case studies: Meta-prototype~\cite{Meta_prototype}, TPU~\cite{TPU}, Edge TPU~\cite{Edge_TPU}, Ascend~\cite{Ascend}, and Tesla NPU~\cite{TeslaNPU}. 
To make a fair and relevant comparison, we normalized all of them to have 1024 MACs and maximally 2MB global buffer (GB) but kept their spatial unrolling and local buffer settings (Table~\ref{table:CS_setting}(a) Idx 1/3/5/7/9). 
Besides, under the concern that all these architectures were originally designed for SL/LBL processing and it thus may or may not be very beneficial to apply DF schedules on them, we manually constructed a DF-friendly versions of all architectures, denoted with `DF' in the end of the name (Table~\ref{table:CS_setting}(a) Idx 2/4/6/8/10).
The guidelines that were followed to construct a DF-friendly version from a SL/LBL architecture are: 1) spatial unrolling is unchanged; 2) the total on-chip memory capacity is unchanged; 3) Input and Output activation are preferably shared in a lower level memory; and 4) Weights should have an on-chip global buffer. 
These guidelines are heuristic-based, and we leave the DF HW architecture optimization problem to future work. 

\begin{figure}[t]
	\vspace{-1em}
	\centering
	\subfloat[Energy (mJ) colorbar]{
		\includegraphics[width=0.215\textwidth,trim={0 0 0 .42cm}, clip]{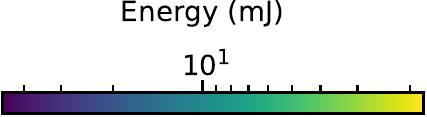}
		\label{subfig:False_False_en}
	}\hspace{2.5mm}
	\subfloat[Latency (million cycles) colorbar]{
		\includegraphics[width=0.218\textwidth,trim={0 0 0 .42cm}, clip]{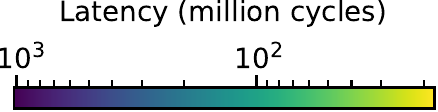}
		\label{subfig:True_False_en}
	} \par
	\subfloat[Fully-recompute, Energy]{
		\includegraphics[width=0.23\textwidth]{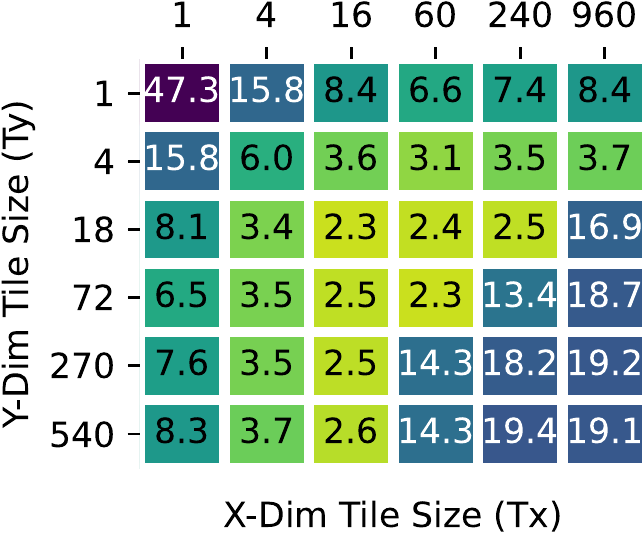}
		\label{subfig:False_False_en}
	}
	\subfloat[Fully-recompute, Latency]{
		\includegraphics[width=0.23\textwidth]{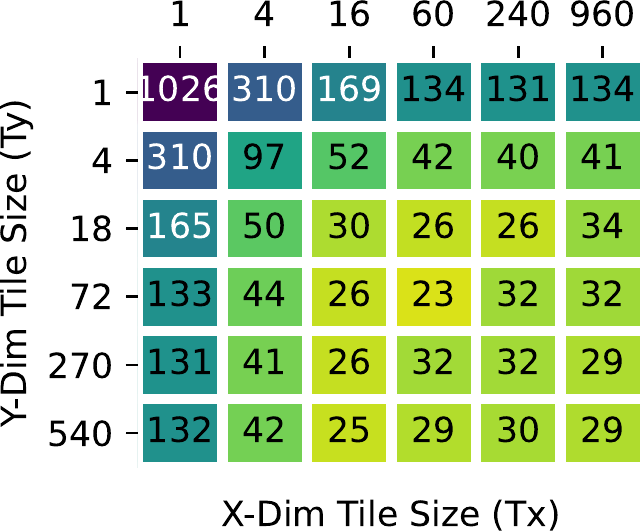}
		\label{subfig:True_False_en}
	} \par
	\subfloat[H-cached V-recompute, Energy]{
		\includegraphics[width=0.23\textwidth]{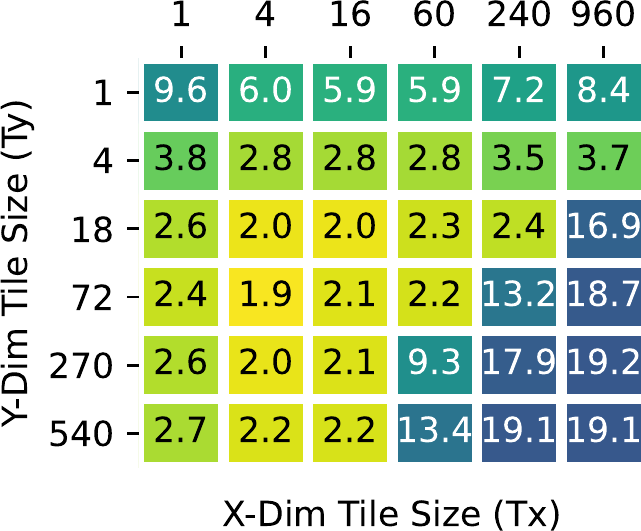}
		\label{subfig:False_False_la}
	}
	\subfloat[H-cached V-recompute, Latency]{
		\includegraphics[width=0.23\textwidth]{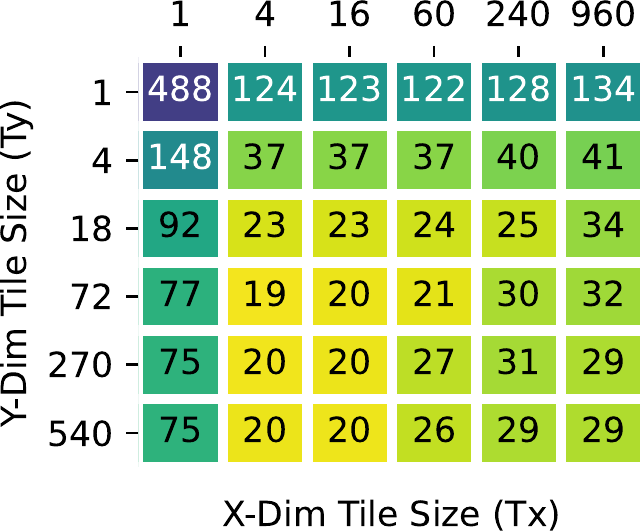}
		\label{subfig:True_False_la}
	} \par
	\subfloat[Fully-cached, Energy]{
		\includegraphics[width=0.23\textwidth]{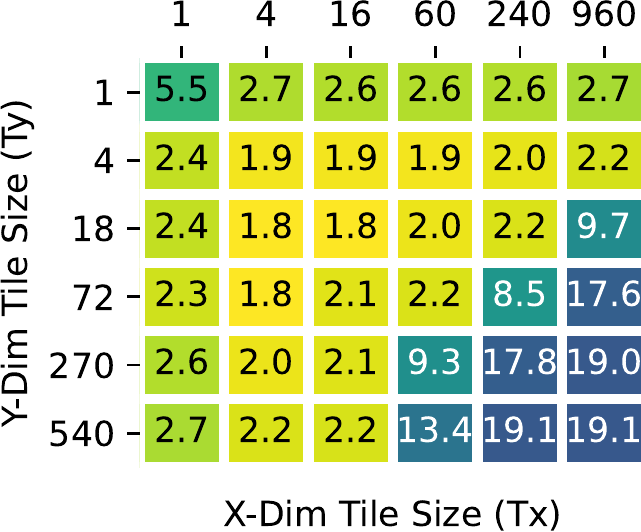}
		\label{subfig:False_False_la}
	}
	\subfloat[Fully-cached, Latency]{
		\includegraphics[width=0.23\textwidth]{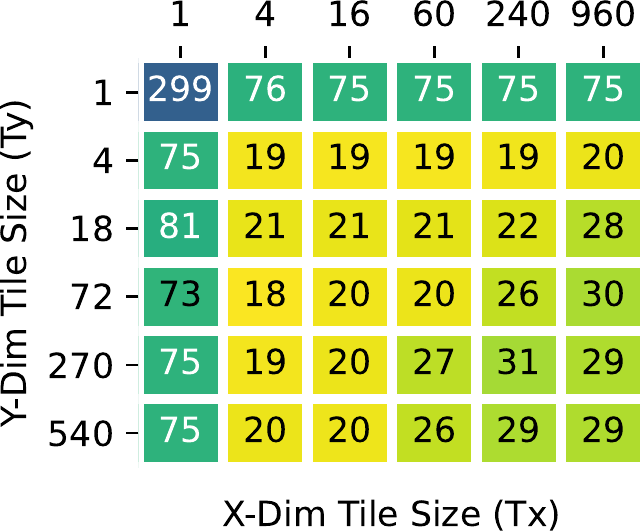}
		\label{subfig:True_False_la}
	}
	\caption{The total energy and latency for meta-proto-like DF architecture processing FSRCNN with different DF strategies.}
\vspace{-1em}
	\label{fig:cs1_heatmap}
\end{figure}

\begin{figure}[!t]
	\centering
	\vspace{-0.5em}
	\includegraphics[width=3in]{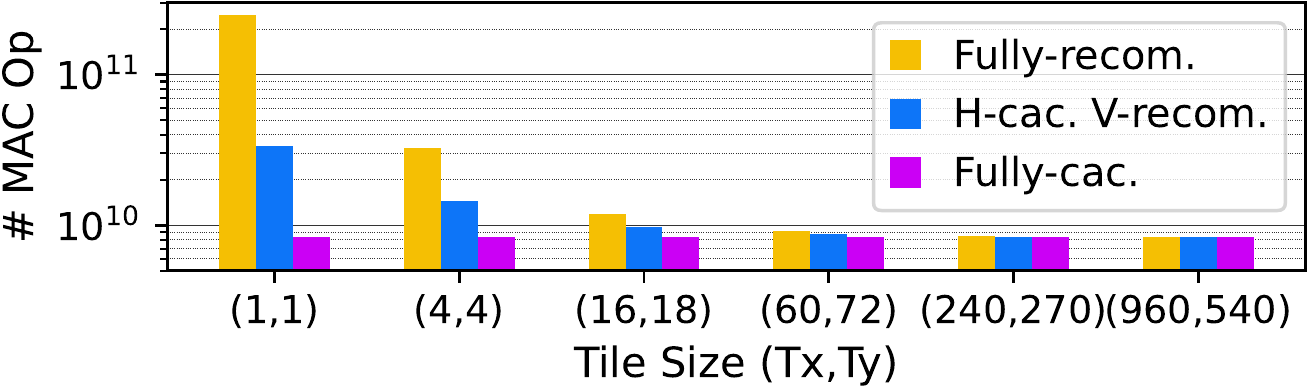}
	\vspace{-0.5em}
	\caption{MAC operation count for different DF strategies.}
	\label{fig:cs1_MAC}
	\vspace{-1em}
\end{figure}

For HW modeling, CACTI7~\cite{CACTI7} is used to extract all the SRAM costs (pJ/word access). Other HW costs, such as Unit MAC, register, and DRAM access cost are scaled accordingly based on the SRAM cost, following the scaling factors reported in \cite{Interstellar}. All the on-chip memory's banking and bandwidth (bit/cycle) are selected in such a way that PE array can get enough data to work at its full speed for ideal workload, while the DRAM bandwidth is fixed to 64bit/cycle to mimic the on-off-chip communication bottleneck.

For workload, five DNN workloads are used in the case studies: FSRCNN~\cite{FSRCNN}, DMCNN-VD~\cite{DMCNN-VD}, MCCNN~\cite{mccnn}, MobileNetV1~\cite{MobileNets} and ResNet18~\cite{ResNet18}. 
Table~\ref{table:CS_setting}(b) shows that FSRCNN, DMCNN-VD, and MCCNN are activation-dominant (all the layers have large feature maps), whereas MobileNetV1 and ResNet18 are weight-dominant (feature maps are smaller and gradually decrease across layers).

Note that the HW architectures picked for these case studies are not DF-specific. 
This enables exploring whether or not these non-DF-specific HW architectures (and their variants with some memory size/sharing adjusting) can benefit from DF scheduling on both activation- and weight-dominant DNNs. 

Another thing to point out is that in DeFiNES, users can self-define the optimizing target (energy, latency, EDP, any memory access, a combination of them, etc.). For the case studies, we prioritized energy.

\begin{figure*}[t]
	\centering
	\subfloat[Layer's Activation (I, O)]{
		\includegraphics[width=0.235\textwidth]{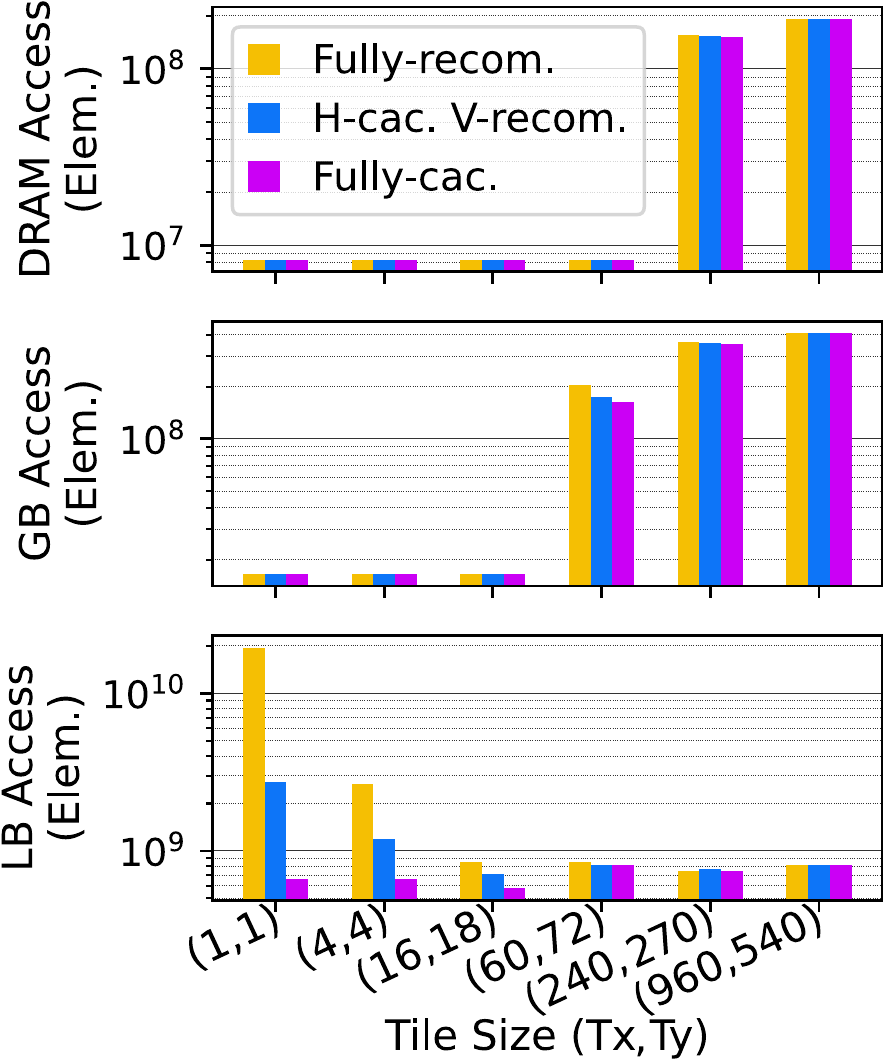}
		\label{subfig:False_False_en}
	}
	\subfloat[Layer's Weight]{
		\includegraphics[width=0.235\textwidth]{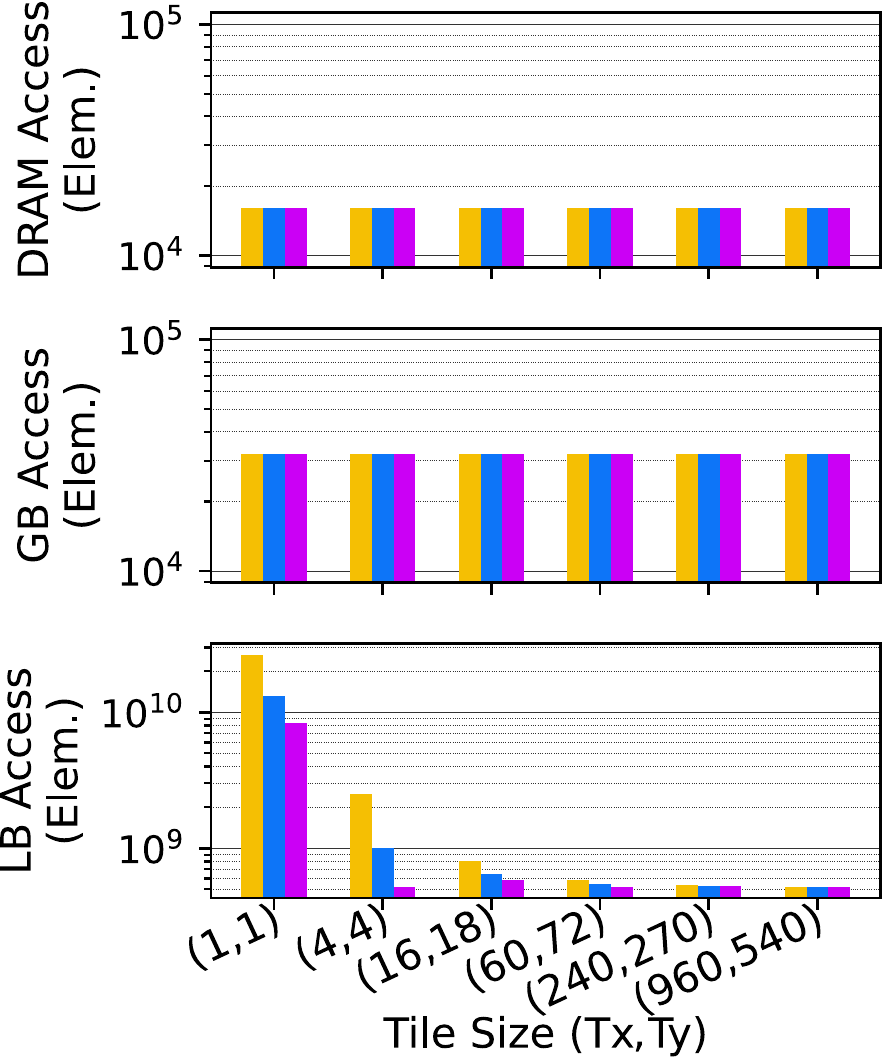}
		\label{subfig:True_False_en}
	}
	\subfloat[Data copy action]{
		\includegraphics[width=0.235\textwidth]{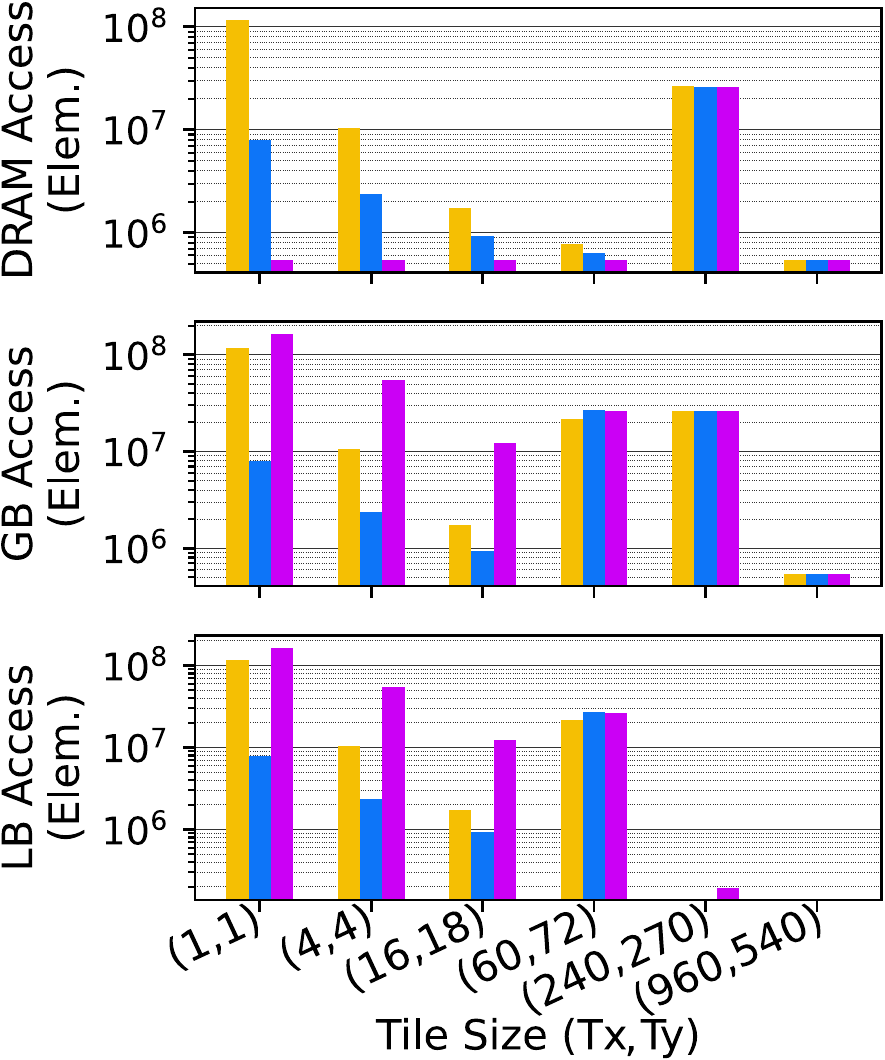}
		\label{subfig:True_True_en}
	}
	\subfloat[Total memory access = (a)+(b)+(c)]{
		\includegraphics[width=0.235\textwidth]{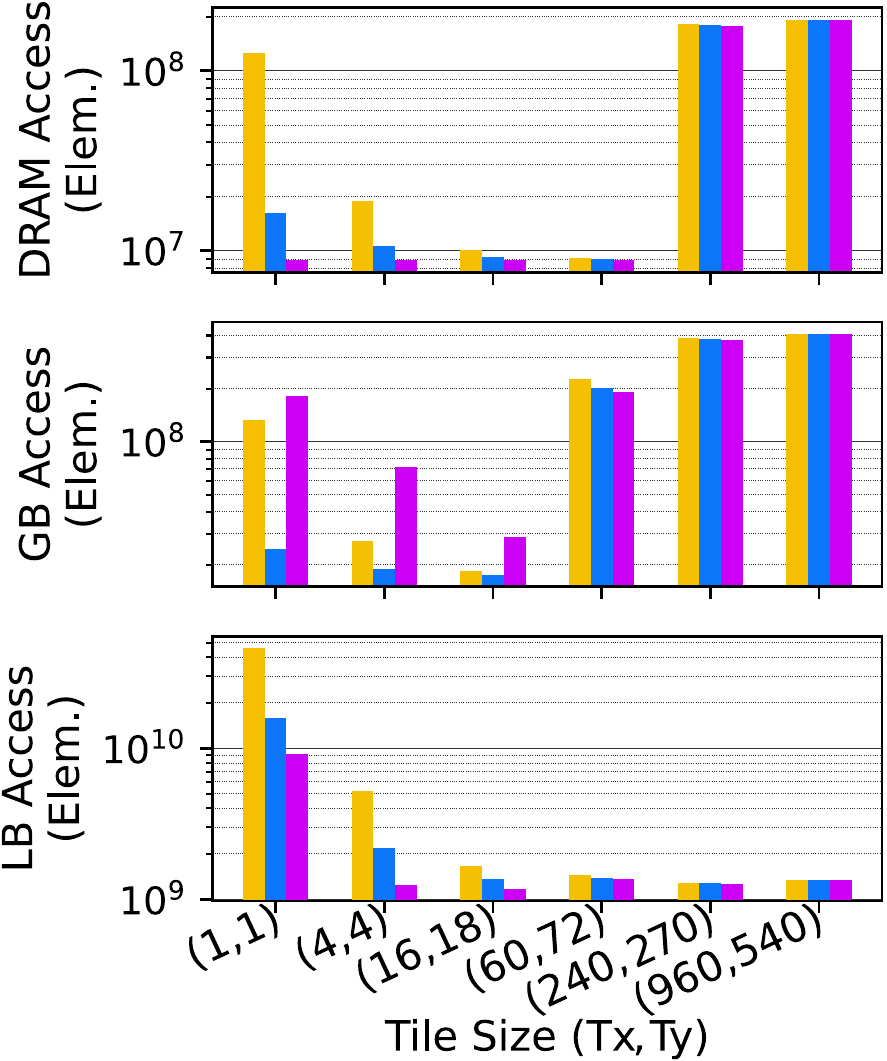}
		\label{subfig:False_False_la}
	}
	\caption{Memory access of different data types at different memory levels for meta-proto-like DF architecture processing FSRCNN with different DF strategies.}
	\vspace{-0.5em}
	\label{fig:cs1_bar_chart}
\end{figure*}

\subsection{Case Study 1: Impact of Depth-first strategy \PageLimit{(1.5)}}\label{sec:cs1}

This case study discusses how much DF strategies impact results when mapping a DNN onto an accelerator, exemplified by
FSRCNN and Meta-proto-like DF (2 in Table~\ref{table:CS_setting}(a)) as the targeted workload and HW architecture.

For the DF scheduling space's three axes (tile size, overlap storing mode, and fuse depth), this case study focuses on exploring the first two axes. The third axis, fuse depth, is fixed to the whole DNN since the total weight size of FSRCNN is small (15.6KB as shown in Table~\ref{table:CS_setting}(b) Idx 1) and thus all weights fit in Meta-proto-like DF architecture's weight on-chip local buffer (32KB as shown in Table~\ref{table:CS_setting}(a) Idx 2). So, there is no benefit to not fuse the whole DNN into one stack, according to the trade-off introduced in Fig.~\ref{fig:ST}.

\begin{figure}[t]
	\centering
	\includegraphics[width=3in]{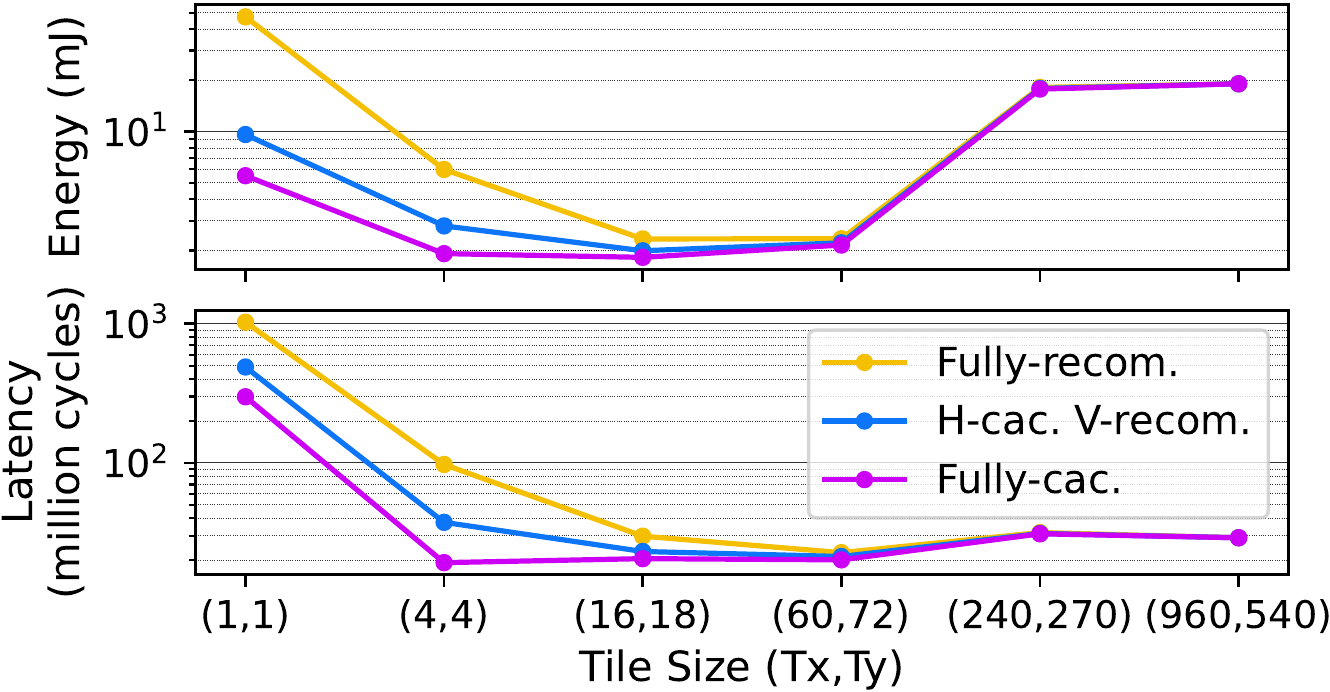}
	\vspace{-0.5em}
	\caption{The total energy and latency for design points in Fig.~\ref{fig:cs1_bar_chart}.}
	\label{fig:cs1_3}
	\vspace{-1em}
\end{figure}

For the first two axes, we swept 110 tile sizes (different spatial dimension tile size (Tx,Ty) combinations) for each of the three overlap storing modes. A subset of the results is shown in Fig.~\ref{fig:cs1_heatmap}, in which the total energy and latency of three overlap
storing modes with different tile sizes are visualized. Note that the figure map size of the last layer of FSRCNN is 960$\times$540, thus all the bottom right blocks in each heatmap (with Tx=960 and Ty=540) correspond to LBL processing. Their energy and latency numbers (19.1 and 29 resp.) are the same because different overlap storing modes do not make a difference for LBL, as discussed in Section~\ref{sec:designspace}.

The rest of this subsection firstly summarizes the main messages delivered by Fig.~\ref{fig:cs1_heatmap}, and then uncovers the causes by using the memory access breakdown of the different types of data of Fig.~\ref{fig:cs1_bar_chart}.

Four major observations can be extracted from Fig.~\ref{fig:cs1_heatmap}: 1) Considering different tile sizes under the same overlap storing mode, \rv{both too small and too large tile sizes are sub-optimal}. The best point is always somewhere in the middle. 2) Considering the same tile size across different overlap storing modes, the order of energy consumption is for most cases: fully-cached $<$ H-cached V-recompute $<$ fully-recompute. 3) Different tile sizes and modes heavily impact energy and latency (up to 26$\times$ difference for energy and 57$\times$ for latency). 4) Fully-recompute prefers larger tile sizes than fully-cached.

To understand the reasons behind, Fig.~\ref{fig:cs1_MAC} and Fig.~\ref{fig:cs1_bar_chart} take out all the diagonal scheduling points from Fig.~\ref{fig:cs1_heatmap}, and respectively plot their MAC operation count and memory access count (in number of data element) for each memory level in the hierarchy (LB, GB, and DRAM) that is contributed by layers' activation, weight and data copy action. Fig.~\ref{fig:cs1_3} further shows the total energy and latency of these diagonal scheduling points.

For layers' activation, Fig.~\ref{fig:cs1_bar_chart}(a) presents two clear trends. Firstly, DRAM and GB access do not depend much on the used mode. When the tile size is small, like (1,1), (4,4) or (16,18), there is little GB and LB memory access because all the activations per tile can fit into LB. When the tile size is increased to a certain point, like (60,72), the GB access suddenly increases due to activations no longer fitting in LB and thus GB being the top activation memory level. Further increasing the tile size till reaches LBL (960, 540) and the DRAM access catches up as a consequence of the intermediate activations no longer being able to fit on-chip. 
Secondly, LB access is very sensitive to the used mode for small tile sizes, with the order of access always: fully-recompute $>$ H-cached V-recompute $>$ fully-cached. This is because more MAC operations are performed when doing re-computation, especially in small tile sizes, as shown in Fig.~\ref{fig:cs1_MAC}, which requires more LB access.

For layers' weight, Fig.~\ref{fig:cs1_bar_chart}(b) shows that all the tile sizes have the same DRAM and GB access, which is reasonable because all the weights of FSRCNN can fit into weight LB. However, for the fully-cached mode, LB weight access is much higher for (1,1) than all other tile sizes.
This is because the spatial unrolling of the HW architecture includes OX 4 $|$ OY 4 (Table \ref{table:CS_setting}(a) Idx 2), and thus tile size (1,1) causes a severe under-utilization of the PE array. This in turn reduces the spatial data reuse of the weight's LB.
In other words, when the tile size (Tx,Ty) $\ge$ (4,4), the spatial unrolling OX 4 $|$ OY 4 can be fulfilled and thus one data read out from weight LB can serve 16 MACs.
In contrast, it can only serve 1 MAC unit per access when the tile size is (1,1).
For the other modes, it is high mainly for the same reason as for activations: there is a relatively large recompute overhead.

Fig.~\ref{fig:cs1_bar_chart}(c) uncovers memory access contributed by data copy actions. As discussed in Section~\ref{sec:model}, data copy actions happen when the required input data of current tile are not all in its lowest-fitting memory level, which could be because the previous layer's output and/or the cached data for reuse have a different lowest-fitting memory level. With this in mind, Fig.~\ref{fig:cs1_bar_chart}(c) is explainable. Firstly, for small tile sizes ((1,1)-(16,18)): 1) fully-recompute mode has large memory access at all memory levels due to the large overlap re-fetching across different tiles of the first DNN layer; 2) fully-cached mode has large memory access at GB and LB memory levels due to the cached data being located in GB while the input's lowest-fitting memory level is LB. Secondly, in middle tile size region ((60,72)-(240,270)), different modes' behaviors converge and the data copy actions mainly come from moving previous layer outputs down to the top memory level of the next layer's input. Lastly, in large tile region, no data copy action is needed as all input, output, and cached data are located in DRAM.

Fig.~\ref{fig:cs1_bar_chart}(d) shows the total memory access, and Fig.~\ref{fig:cs1_3} visualizes the overall energy and latency, which together with the memory access breakdown discussed earlier help us to better understand the heatmaps in Fig.~\ref{fig:cs1_heatmap}:  for fully-recompute mode, the small tile sizes' sub-optimality comes from data re-fetching and MAC re-computation of the large overlap region; for fully-cached mode, the small tile sizes' sub-optimality comes from the large weight access and cached data movement; for all the modes, large tile sizes' sub-optimality is due to large DRAM access of activation.

This case study shows that different DF strategies vary a lot on energy and latency, and DeFiNES can analyze/reason about them, taking the advantages of the unified analytical model.

\subsection{Case Study 2: Applying DF to Multiple Workloads \PageLimit{(0.5)}}\label{subsec:cs2}

This case study studies how different workloads \rv{prefer different} DF strategies.
To this end, we map all five workloads of Table \ref{table:CS_setting}(b) on the meta-proto-like hardware and compare five different inference strategies:

\begin{itemize}
	\item \textbf{Single layer}: layers are completely evaluated one at a time, feature maps are always stored to and fetched from DRAM in between layers;
	\item \textbf{Layer-by-layer}: layers are completely evaluated one at a time, intermediate feature maps are passed on to the next layer in the lowest memory level they fit in;
	\item \textbf{Fully-cached DF with 4$\times$72 tiles}, which is the \rv{best}  found in case study 1;
	\item The best strategy found when a \textbf{single strategy} is used for all fused layer stacks;
	\item The \textbf{best combination}, where different stacks can use different DF strategies.
\end{itemize}

\begin{figure}[t]
	\centering
	\includegraphics[scale=0.9]{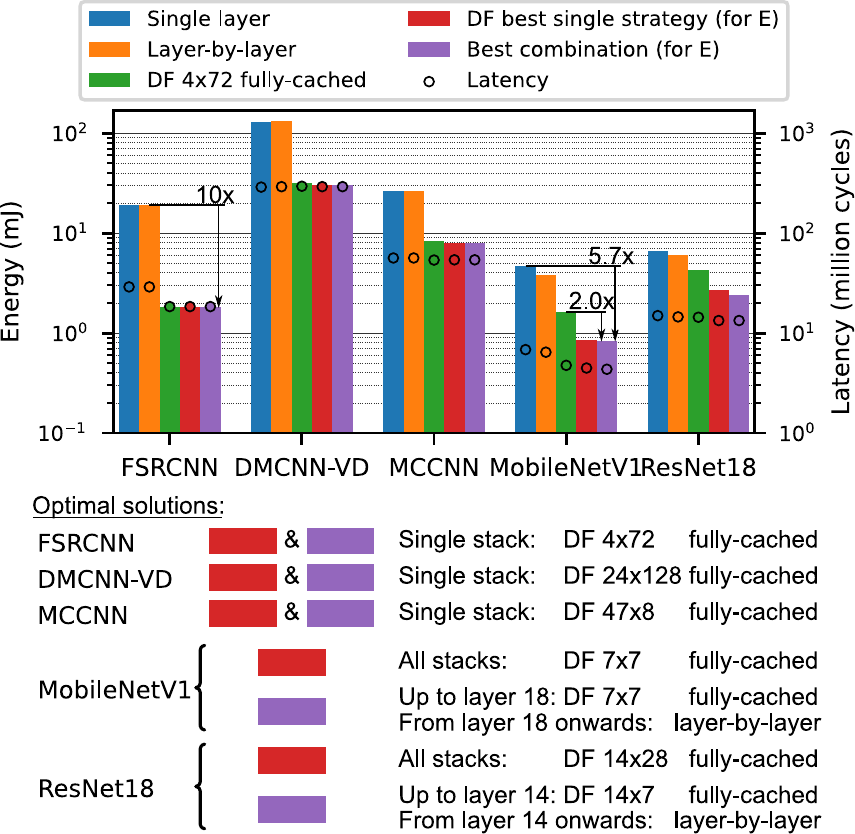}
	\vspace{-0.5em}
	\caption{Case study 2: Different workloads lead to different \rv{best} solutions (all results on meta-proto-like DF hardware)}
	\label{fig:cs2}
	\vspace{-1.5em}
\end{figure}

Fig. \ref{fig:cs2} visualizes the results, which show some noteworthy findings.
Firstly, for the workloads with spatially large features maps (FSRCNN, DMCNN-VD and MCCNN), their individual \rv{best} solutions (purple) are not significantly better than the \rv{best} solution found in case study 1 (green). 
The latter is thus a very good solution across a range of workloads similar to the one it was found for, with a gain of $10\times$ compared to SL.

Secondly, this solution does not perform as well on MobileNetV1 and ResNet18, which operate on spatially smaller feature maps with more channels. 
On MobileNetV1 for instance, it is $2.0\times$ worse than the \rv{best found} result.
In these workloads, the deeper layers are more weight-dominant, which impedes fusing them into one stack.
Hence, the combined \rv{best} solution applies DF to the first, activation-dominant layers and LBL to the last, weight-dominant layers.
This combination achieves a gain of $5.7\times$ over SL on MobileNetV1.

\subsection{Case Study 3: A Joint DSE of Accelerator Architecture and Scheduling for Multiple Workloads \PageLimit{(0.5)}}\label{subsec:cs3}

This case study examines the effect of the accelerator's architecture on the optimal inference strategy.
In particular, it compares the default accelerators architectures, which were designed with LBL inference in mind, against the manually adjusted DF-friendly variants by looking at the geometric average of performance across all five workloads of Table \ref{table:CS_setting}, with both LBL and DF best single strategy (for energy).

\begin{figure}
	\centering
	\includegraphics[width=\columnwidth]{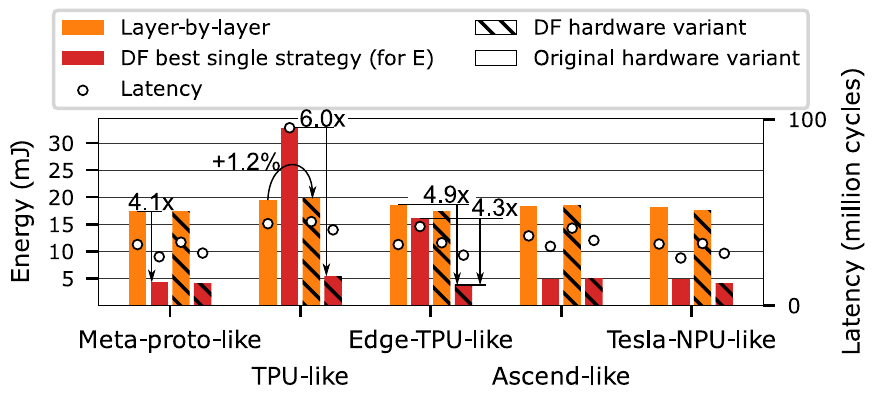}
	\vspace{-1.5em}
	\caption{Case study 3: Different HW architectures' energy and latency (geometric mean across the 5 workloads) when applying layer-by-layer or best DF scheduling strategies.}
	\vspace{-1em}
	\label{fig:cs3}
\end{figure}

The results in Fig. \ref{fig:cs3} 
show that DF outperforms LBL on all accelerator architectures except for TPU-like, including the unadjusted default accelerators, on which the maximum gain was $4.1\times$.
TPU-like has poor support for DF schedules due to the absence of on-chip weight buffers.
With such a buffer added in the DF-friendly variant, DF significantly outperforms LBL, 
indicating the importance of designing with DF compatibility in mind. This finding is further backed by the overall comparison between the DF-friendly and default variants, which
shows that the DF-friendly variants are at least as good as the defaults when using DF, with large gains of $6.0\times$ and $4.3\times$ for TPU-like and Edge-TPU-like hardware respectively, and maximally 1.2\% worse when using LBL.

Overall the biggest difference (in geometric mean over the five workloads) between LBL inference on default hardware variants and DF on DF-friendly variants is found for the Edge-TPU-like hardware and equals $4.9\times$.

\section{Related Works \PageLimit{(0.75)}} \label{sec:sota}
Previous works on DNN DF processing can be split into two categories, DF-supporting HW implementations and DF modeling and exploration frameworks.

For DF-supporting HW implementations, designs\cite{15FusedLayerCNN_2016MICRO,18AFullHD60_2019VLSI,22DaduEye_2021JSSCC,23A121T_2022JSSCC,24DTCNN_2020TCASI,DepFiN_2021VLSI,17SmartphoneSoC_2020ISSCC,16ECNN_2019MICRO} have demonstrated their DF solutions and shown large benefits in terms of energy and/or latency compared to the traditional single-layer/layer-by-layer DNN accelerators for targeted workloads.
These DF-supporting designs each have a default depth-first processing pattern, with manually selected tile size, overlap storing mode, and fuse depth. 
For example, regarding the overlap storing mode, \cite{16ECNN_2019MICRO} fully recomputed the intermediate overlapping data while \cite{18AFullHD60_2019VLSI} assumed a horizontally-cached with vertical recompute mode and \cite{DepFiN_2021VLSI} applied a fully-cached mode.
Regarding the depth of the stack of fused layers, \cite{17SmartphoneSoC_2020ISSCC} chose to fuse just 2 layers at a time while \cite{18AFullHD60_2019VLSI, DepFiN_2021VLSI} preferred deeper fused stack (8-20 layers).
On the tile size, \cite{22DaduEye_2021JSSCC,23A121T_2022JSSCC} always treated one row of an image as a tile, whereas \cite{15FusedLayerCNN_2016MICRO,24DTCNN_2020TCASI} adopted a square tile size and \cite{DepFiN_2021VLSI} has a preferred tile size of 128 pixels along an image row. 
All these HW implementations are optimized for one or a few types of predefined DF strategies, and it is unclear if there are remaining combinations of HW architectures and DF strategies that would perform better for a targeted workload.
Researching this would preferably be done at a high abstraction level to save the time of fully simulating and/or developing the HW architectures.

Therefore, %
%
several DF modeling and exploration frameworks like DNNVM \cite{DNNVM}, Efficient-S \cite{EfficientS}, LBDF \cite{LBDF}, ConvFusion \cite{ConvFusion}, Optimus \cite{Optimus}, and DNNFuser \cite{DNNFuser} have been proposed. 
These frameworks, listed in Table \ref{table:related_work}, help to model and optimize the DF schedule given HW architectures and DNN workloads. 
In the optimizing part, many innovative searching algorithms are introduced, such as the heuristic subgraph isomorphism algorithm in DNNVM, the DAG-based hardware-aware operator fusion algorithm in Optimus, and a transformer-based mapper in DNNFuser.
However, they all have some important \rv{factors} missing in the modeling part. 

\begin{figure}[tb]
	\centering
	\includegraphics[width=3.3in]{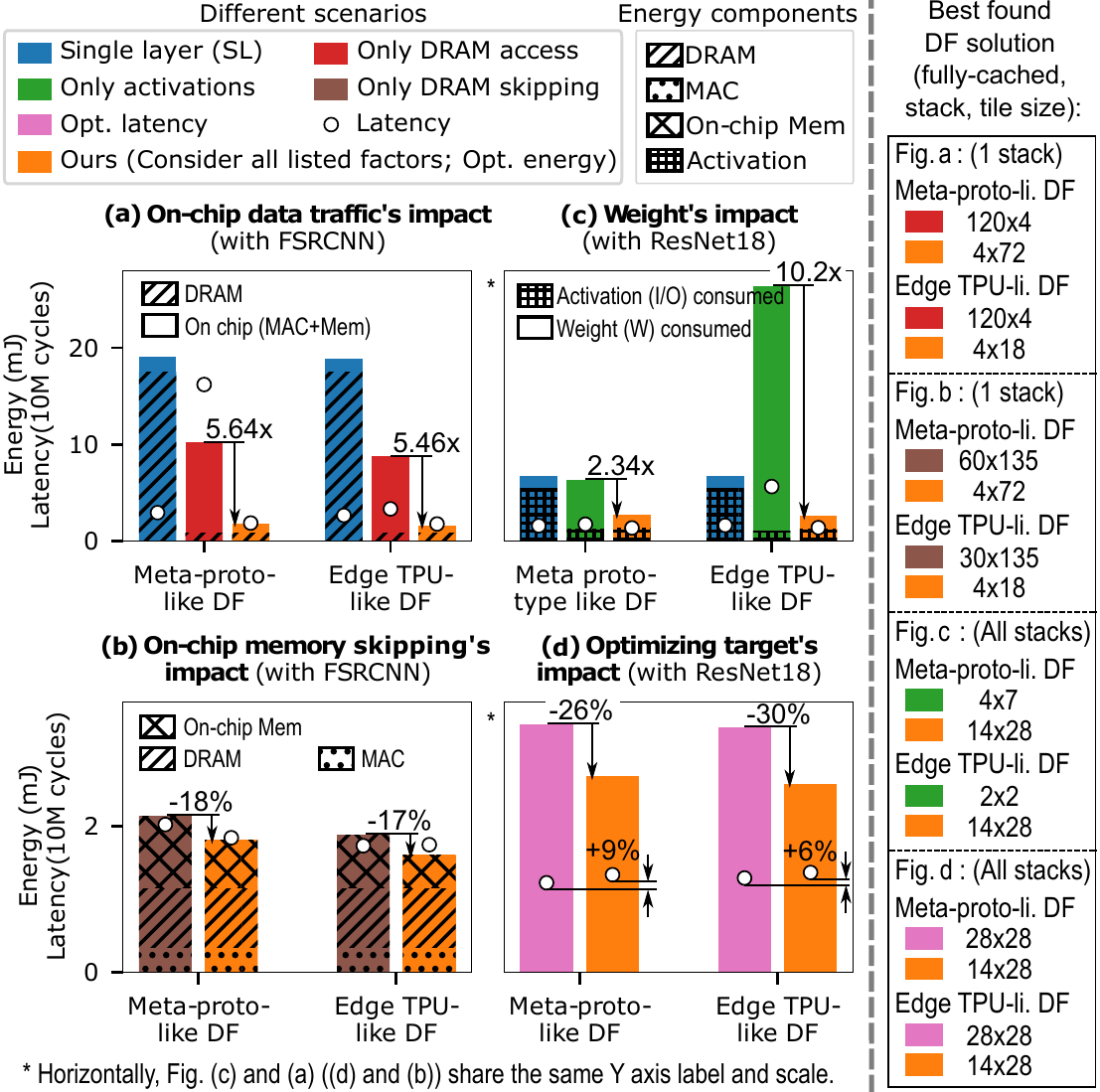}
	\vspace{-0.5em}
	\caption{\rv{Experiments to evaluate different factors in Table~\ref{table:related_work}.}}
	\label{fig:sota}
	\vspace{-1.5em}
\end{figure}

\rv{The rest of this section will discuss each missing factor (as in Table \ref{table:related_work}), and some of theirs impact (as in Fig.~\ref{fig:sota}).}

Firstly, from the DF scheduling space point of view, most of these frameworks do not explicitly support exploring the trade-offs between different overlap storing modes. As shown in \rv{Fig.~\ref{fig:tile_type_count}} and Section~\ref{sec:cs1} (Case Study 1), this can have a big impact on \rv{tile type count (related to code and control
complexity)} and system energy/latency.

Secondly, from the HW modeling point of view, they only focus on \rv{modeling/optimizing the DRAM access} while ignoring the data movement within the potential multi-level on-chip memory hierarchy. 
In other words, they are agnostic of on-chip memory hierarchy. 
This could cause substantial loss\rv{es}, as proven by \rv{Fig.~\ref{fig:sota}(a)}, which shows the experiment results of mapping FSRCNN onto two HW platforms in three ways: 1) Single-Layer (SL), 2) DF but only optimize for DRAM traffic, 3) DF and optimized for the overall energy (our work). 
The DRAM energy contribution is highlighted by the diagonal hatching, which shows that DRAM energy dominates in the SL case.
Using DF and only optimizing for DRAM traffic, the DRAM energy can indeed be largely reduced, but omitting the on-chip energy (non-hatched part in the red bar) from the optimization can make the latter dominant. Only when considering the whole system, the \rv{best} DF solutions (orange bars) can be achieved. 
The parameters of the found solutions (Fig.~\ref{fig:sota} \rv{right}) show that when optimizing for the overall energy (orange), the framework found a smaller tile size compared to optimizing for DRAM only (red). 
This can be explained: \rv{1)} When optimizing for DRAM only, the tool will \rv{randomly pick one DF schedule that makes sure all the intermediate data fit on chip, and thus DRAM access is minimized. However, after achieving the minimal DRAM access, there is still a lot of room for on-chip data traffic optimization, which is overlooked in this case.}
\rv{2)} When optimizing for the overall energy, it benefits from smaller tile sizes since at a certain point, not only can all the data of intermediate tiles fit in on-chip GB, but also fit in the LB. \rv{In this case, the activation can be fully reused in LB, and GB access is minimized (on top of the already minimized DRAM access), resulting in a $5.64\times$ energy gain for FSRCNN on the meta-proto-like DF HW.}

\rv{Thirdly, on top of modeling on-chip data traffic, we further evaluated the benefit of performing multi-level memory skipping over DRAM-only skipping, i.e. skipping (multiple)  upper (on-chip) memory level(s) when writing back the outputs of intermediate tiles if it they fully fit in lower level memories. Around 17\%-18\% energy gain is observed for the tested workload-HW combination, as shown in Fig.~\ref{fig:sota}(b). Due to this step targeting optimizing on-chip memory energy, the gain is not very significant if the MAC energy and the (already minimized) DRAM energy are dominant, which is the case here. This technique can bring larger gains for systems with more dominant on-chip data traffic.} 


\begin{table}[t]
	\centering
	\caption{\rv{Related DF modeling framework comparison}}
	\vspace{-0.5em}
	\includegraphics[width=3.25in]{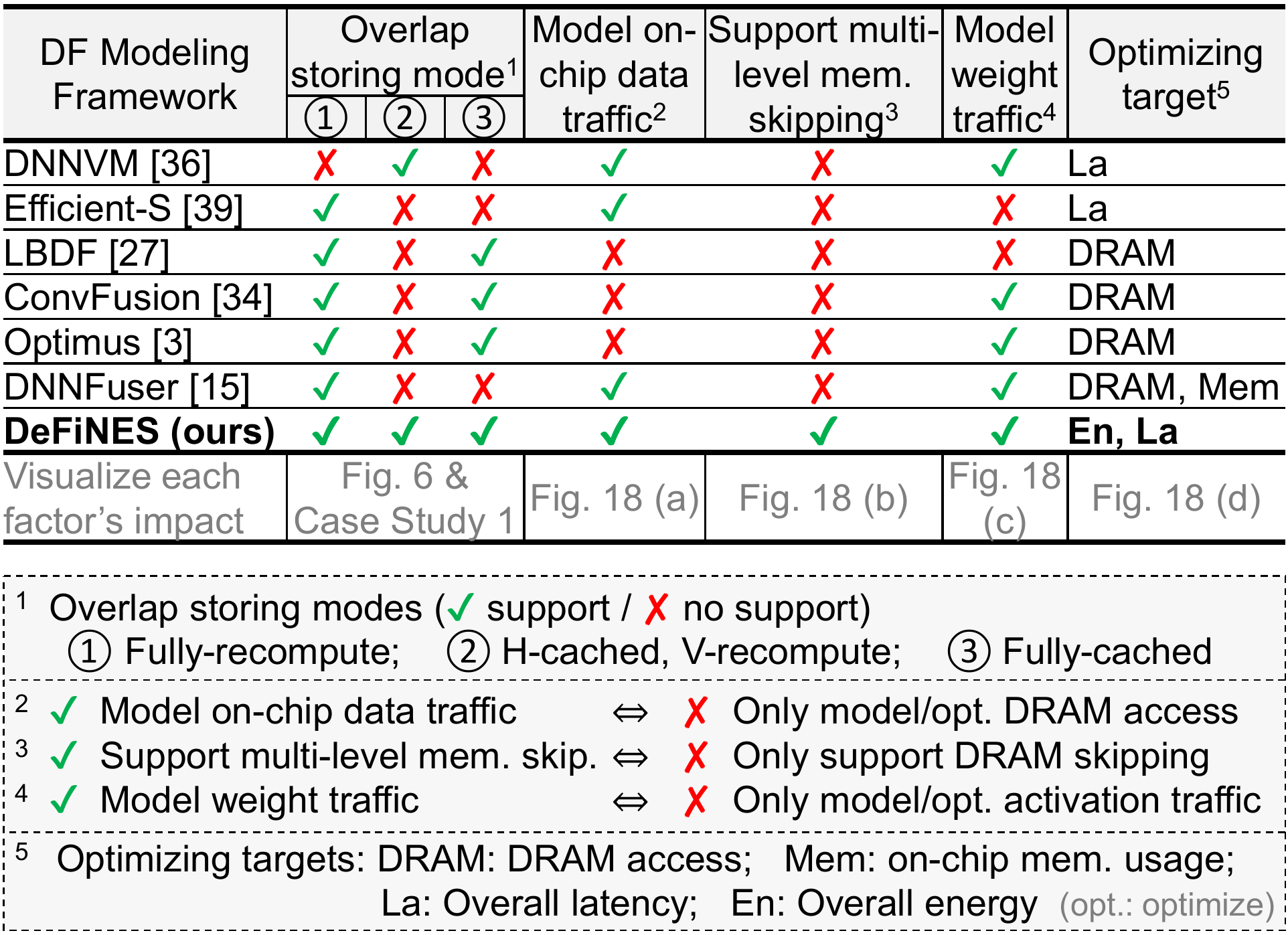}
	\label{table:related_work}
	\vspace{-1.5em}
\end{table}%

\rv{Fourthly}, most of DF HW implementations and exploration frameworks show the energy, latency, and/or DRAM access gain that come from activation tiling, but do not mention much about the potentially higher weight energy costs due to the loss of local weight data reuse.
This can be harmful for the overall system efficiency, as shown by the example of \rv{Fig.~\ref{fig:sota}(c)}. \rv{The energy portion caused by memory access for activations, highlighted with square hatching, contributes most of the energy in the SL case.}
However, just blindly optimizing for activations while ignoring the weights ends up in the green bars.
While these indeed have minimal energy caused by activations, the energy caused by weights' memory accesses dominates and causes a large penalty (non-hatched part in the green bars). 
This is because the tool found very small tile sizes as its \rv{best} solution when only optimizing for activation. 
This lets activations skip higher level memories as much as possible, but at the same time largely reduces the low-level memory's weight data reuse, thus triggering more access to higher level weight memories.
So, only when considering both the benefit and drawbacks that tiling can bring, the \rv{best} DF solution (orange bars) can be achieved.
For the given example, taking weights into account achieves a solution that has $2.34\times$ and $10.2\times$ less energy than the solution found by only considering activations for the meta-proto-like DF and Edge-TPU-like DF hardware architectures, respectively.


\rv{Lastly, different frameworks have different optimizing targets, as shown in the last column of Table~\ref{table:related_work}: some of the frameworks only evaluate latency while ignoring energy, whereas some only care about optimizing the DRAM access. As DRAM-only optimization's downsides have been explained, here we focus on discussing latency- and energy-optimized solution comparison. Fig.~\ref{fig:sota}(d) shows the results: pink/orange bars are the energy (and the corresponding dots are latency) of our latency-/energy-optimized DF schedules respectively. In this example, a clear latency-energy trade-off is presented and the best found DF solution shows that the energy-optimized DF schedule prefers a smaller tile size than the latency-optimized one. This is because smaller tile sizes on one hand help reduce energy by enabling skipping of more memory levels while, on the other hand, it increases the data preparation cycle (loading and offloading) overhead.}

To summarize, our work models the complete DF design space with support for detailed activation and weight, on- and off-chip memory hierarchy analysis so as to better capture the trade-offs between different DF strategies \rv{and optimizing targets}.
These properties enable DeFiNES to make \rv{the overall best} choices without neglecting factors that may turn out to be important otherwise.
This makes DeFiNES a good addition to the previously mentioned optimization-oriented frameworks ~\cite{DNNVM,EfficientS,ConvFusion,Optimus,DNNFuser}. Together with those, we can better design and schedule DNN accelerators.


\section{Conclusion\PageLimit{(0.25)}}
This work first presented a definition of the depth-first (a.k.a. layer fusion, or cascaded execution) design space, and then a cost model capable of handling this whole design space.
Furthermore, the cost model considers not only DRAM access or only memory access due to activations, but also the full on-chip memory hierarchy and memory access caused by weight traffic.
Large gains might be missed when not doing so (up to $10.2\times$ in the shown examples; \rv{Fig.~\ref{fig:sota}(c)}).


Using this model, the case studies showed that depth-first strategies can significantly outperform layer-by-layer execution, 
even when the workload is not activation-dominant (MobileNetV1 and ResNet18), and even when the hardware is not designed for it: depth-first strategies outperformed layer-by-layer on four of the five tested hardware architectures with gains of up to $4.1\times$.
However, some architectures may be ill suited for depth-first, in which case small adjustments to their design can lead to large improvements. 
For instance, reassigning some of the on-chip memory capacity of the TPU-like architecture enabled it to greatly benefit from depth-first strategies, outperforming its default variant by $6\times$. 
These examples show how DeFiNES allows to quickly examine the complex design space of different combinations of depth-first strategies and hardware architectures.



\section*{Artifact Appendix}

\subsection{Abstract}

Our artifact provides a guide to replicate the primary experiments (case study 1) demonstrated in this paper. Since case study 2 and 3 are equivalently performing case study 1 multiple times (for different workload and HW architecture combinations) and will take a long time to run and generate multi-Gigabytes of data, we here focus on case study 1.

The included materials are the source code of DeFiNES and the scripts to auto-run the experiments, collect data, and make the plots. In the end, we also provide the useful information on experiment customization, i.e. users can use DeFiNES to carry out their own DNN accelerator-schedule design space exploration, considering both layer-by-layer and depth-first scheduling possibilities.

\subsection{Artifact Check-list (meta-information)}

\begin{itemize}
\item \textbf{Algorithm} DeFiNES
\item \textbf{Program} Python 3.9 program
\item \textbf{Compilation} Normal python compilation
\item \textbf{Run-time environment} 
\\Linux or Windows with Anaconda installed
\item \textbf{Hardware} 
General-purpose computer
\item \textbf{Execution} Run python scripts
\item \textbf{Metrics} 
\\Energy, latency, memory access, MAC count, and so on.
\item \textbf{Output} 
\\Pickle files (.pkl) that include all the result details and PDF files that corresponds to several figures in our paper
\item \textbf{Experiments} 
\\Applying different depth-first scheduling options for processing one neural network workload (FSRCNN) on one DNN accelerator (Meta-proto-like DF architecture)
\item \textbf{How much disk space required (approximately)?}
\\150 MiB to store the artifact directory and the results
\item \textbf{How much time is needed to prepare workflow (approximately)?}
\\20 minutes to install Anaconda
\\3 minutes to settle the python environment in Anaconda
\item \textbf{How much time is needed to complete experiments (approximately)?} 18 hours (with 1 CPU thread)

\item \textbf{Publicly available?}
Yes, archived on Zenodo and open-sourced on GitHub
\item \textbf{Code licenses}
BSD 3-Clause License
\end{itemize}

\subsection{Description}

\subsubsection{How to access}
This artifact version is archived on Zenodo at \url{https://doi.org/10.5281/zenodo.7384293}. The project is also open-source on GitHub at \url{https://github.com/ZigZag-Project/DeFiNES}.

\subsubsection{Hardware dependencies}
Any PC with at least 4GB of RAM

\subsubsection{Software dependencies}
Python 3.9 or higher, numpy, networkx, sympy, matplotlib (We provide a .yml file to settle the conda environment for you)

\subsection{Installation}

\begin{itemize}
\item Install Anaconda from \url{https://www.anaconda.com/}
\item Download DeFiNES from \url{https://github.com/ZigZag-Project/DeFiNES} and \texttt{cd} into the repo
\item Create and activate the environment with the provided environment.yml\\
\texttt{conda env create -f environment.yml}\\
\texttt{conda activate DeFiNESenv}
\end{itemize}

\subsection{Experiment Workflow and Expected Results}
After the previous Installation is done, two steps are required to run the case study 1 and reproduce the overall comparison result (Fig.~\ref{fig:cs1_heatmap}) and the detailed analysis results (Fig.~\ref{fig:cs1_MAC}, Fig.~\ref{fig:cs1_bar_chart}, Fig.~\ref{fig:cs1_3}, Fig.~\ref{fig:tile_type_count}(left), Fig.~\ref{fig:tile_memory_levels} and Fig.~\ref{fig:tile_data_sizes}). 

Note that these two commands need to run in sequence under the repo folder.

\subsubsection{Step 1} Run \texttt{python main\_artifact.py} 
\begin{itemize}
\item \textit{What does this script do? }
\\It applies 108 depth-first scheduling options (3 modes with 6$\times$6 X-Dim and Y-Dim tile size combinations) for processing FSRCNN on Meta-proto-like DF architecture. 
\item \textit{Run time?}
\\18 hours with the default setting (using 1 CPU thread and set \texttt{loma\_lpf\_limit=8}). \texttt{loma\_lpf\_limit} is a speed-quality tradeoff tuning knob. User can change its value in \texttt{main\_artifact.py}. The larger it is, the longer the program runs, and possibly the better the result found. For all the experiments in the paper, we set it to 8 to guarantee the best results can be found. 
\\For testing purpose, users can set \texttt{loma\_lpf\_limit} to 6, the total run time will be reduced dramatically from 18 hours to 45 minutes, while some design points' best found energy will increase by a few percents. So, the figures plotted in this case will be slightly different than the original ones in the paper.
\item \textit{What results are expected?} When the program finishes, an overall energy and latency comparison figure will be plotted for these 108 depth-first scheduling options (Fig.~\ref{fig:cs1_heatmap}), and 108 result pickle files (.pkl) will be saved into the \texttt{result\_saving\_path} defined in the \texttt{main\_artifact.py} (by default, it is \texttt{.\textbackslash result\_pickle\_files\textbackslash}).
\end{itemize}

\subsubsection{Step 2} Run \texttt{python plot\_artifact.py} 
\begin{itemize}
\item \textit{What does this script do?}
\\It extracts the required information from the generated result pickle files and makes the plots.
\item \textit{Run time?} 1 minute
\item \textit{What results are expected?} 
\\ Multiple detailed analysis figures: Fig.~\ref{fig:cs1_MAC}, Fig.~\ref{fig:cs1_bar_chart}, Fig.~\ref{fig:cs1_3}, Fig.~\ref{fig:tile_type_count}(left), Fig.~\ref{fig:tile_memory_levels} and Fig.~\ref{fig:tile_data_sizes}.
\end{itemize}

In the end, all the plots will be saved to \texttt{.\textbackslash result\_plot\textbackslash}.

\subsection{Experiment Customization}
The goal of this work is to provide an open-source framework for DNN accelerator architecture-schedule optimization, which allows users to plug in their own setting files and perform customized design space exploration experiments.

For this, users need to provide DeFiNES with the inputs listed in Fig.~\ref{fig:overview}: a workload, a HW architecture, and some depth-first scheduling parameters (in which the Fuse depth is set automatically by DeFiNES, thus no need to provide).

This work has 5 workloads and 10 HW architectures modelled for the case study 2 and 3. Users can perform experiments on them, modify them to try new design options, or create own workloads and/or HW architectures following the same data formats in these example setting files. More details on how to set the DeFiNES input files are provided on the GitHub page.

An example command users can run (Use \texttt{python main.py --help} to see what each argument means):

\begin{itemize}
\item \texttt{python main.py --accelerator inputs.HW.Edge\_TPU\_like --workload inputs.WL.Edge\_TPU\_like.workload\_mccnn --dfmode 1 --tilex 16 --tiley 8} 
\end{itemize}

The results are saved as pickle files (.pkl) in the pre-defined \texttt{result\_saving\_path}. User can use/modify the functions provided in \texttt{plot\_artifact.py} and \texttt{plot\_helper\_funcs.py} to extract various data from the pickle files and visualize the results.

We are continually improving the framework on GitHub, and welcoming all questions and feedback. We hope our tool can help other researchers to better explore and understand the vast DNN accelerator architecture and scheduling design space and can offer the best design solutions.

\section*{Acknowledgements}
This work was supported by the Flemish Government (AI Research Program), the European Commission through the project CONVOLVE (101070374), and the Reality Labs, Meta.


\bibliographystyle{IEEEtranS}
\bibliography{refs}

\end{document}